\documentclass[12pt,a4paper]{article}
\usepackage{amsmath,amsfonts,amsthm,graphicx,lscape}
\usepackage[dvips]{psfrag}
\usepackage{cite}

\usepackage{textcomp}
\usepackage{amsmath,amsthm,amssymb}
\usepackage{amsfonts,graphicx,lscape}

\usepackage{accents}
\makeatletter
\def\widebar{\accentset{{\cc@style\underline{\mskip10mu}}}}
\makeatother

\newcommand{\vt}[1]{\mbox{\boldmath$#1$}}

\providecommand{\abs}[1]{\lvert#1\rvert}

\numberwithin{equation}{section}

\begin{document}
\title{
\vspace{-10mm}
On a \mbox{$(2+1)$}-dimensional generalization 
of the Ablowitz--Ladik lattice 
and 
a discrete 
Davey--Stewartson system
}
\author{Takayuki \textsc{Tsuchida}\footnote{
E-mail:\ 
tsuchida at 
poisson.ms.u-tokyo.ac.jp
}
\vspace{1.5mm}
\\
{\it Okayama Institute for Quantum Physics,
}
\\
{\it Kyoyama 1-9-1, Okayama 700-0015, Japan}
\\
\\
Aristophanes \textsc{Dimakis}\footnote{
E-mail:\ dimakis at aegean.gr}
\vspace{1.5mm}
\\
{\it Department of Financial and Management Engineering}
\\
{\it University of the Aegean, 41 Kountourioti Str.,} 
\\
{\it GR-82100 Chios, Greece}
}
\maketitle
\begin{abstract}
We propose a 
natural 
\mbox{$(2+1)$}-dimensional generalization of 
the Ablowitz--Ladik lattice that is 
an integrable space discretization of the 
cubic 
nonlinear Schr\"odinger (NLS) system 
in \mbox{$1+1$} dimensions. 
By further 
requiring 
rotational symmetry of order $2$ 
in the 
two-dimensional 
lattice, 
we identify an appropriate change of dependent variables, which
translates
the \mbox{$(2+1)$}-dimensional Ablowitz--Ladik lattice 
into a 
suitable space discretization of the Davey--Stewartson system. 
The 
space-discrete Davey--Stewartson system 
has 
a 
Lax pair 
and 
allows 
the complex conjugation reduction 
between two dependent 
variables as in the continuous case. 
Moreover, it is 
ideally 
symmetric 
with respect to space reflections. 
%
Using 
the 
Hirota
bilinear
method, 
we construct 
some exact solutions such as 
multidromion solutions. 
%
%
\end{abstract}
\vspace{5mm}

%
\newpage
\noindent
\tableofcontents

\newpage
\section{Introduction}

More than 40
years have passed since 
the Korteweg--de Vries (KdV) 
equation 
was 
solved 
by Gardner {\it et al.}~\cite{GGKM}
using the inverse scattering method
based on 
its Lax pair~\cite{Lax}. 
The number of 
known integrable 
systems 
following 
the KdV equation, 
particularly 
partial differential equations (PDEs)
in \mbox{$1+1$} space-time dimensions, 
has 
increased 
enormously, 
and 
various techniques to 
study them 
have been 
developed. 
Recently the center of researchers' interest 
has 
shifted from 
continuous 
PDEs 
to 
differential-difference or partial difference equations
wherein at least one of the independent variables 
takes discrete values.
A major problem in 
this 
trend 
is 
how to find a suitable 
difference 
analogue of a 
given 
differential equation. 
The suitable discretization 
of an integrable continuous system 
is generally 
required to retain 
the integrability~\cite{Suris03}, 
but 
that is not sufficient if 
the original 
continuous 
system has 
some essential 
internal symmetries. 
This becomes conspicuous if we 
consider 
integrable discretizations of 
nonlinear Schr\"odinger (NLS)-type systems, 
i.e., 
two-component 
systems of second order 
that allow 
the complex conjugation reduction 
between the two dependent variables. 
As a prototypical example, we 
discuss 
the 
cubic NLS system in \mbox{$1+1$} dimensions~\cite{AKNS}, 
\begin{subequations}
\label{nonredNLS}
\begin{align}
& \mathrm{i} q_t + q_{xx} - 2q^2 r = 0,
\label{NLS1}
\\
& \mathrm{i} r_t - r_{xx} + 2r^2 q = 0.
\label{NLS2}
\end{align}
\end{subequations}
Note that 
the reduction 
\mbox{$r=\sigma q^\ast$}
with a real constant $\sigma$ 
simplifies 
the two-component system 
(\ref{nonredNLS}) to 
the scalar NLS equation~\cite{ZS1,ZS2}. 
In addition, 
(\ref{nonredNLS}) is invariant under the space reflection 
\mbox{$x \to -x$} 
as well as 
the time reflection 
\mbox{$t \to -t$} with \mbox{$q \leftrightarrow r$}. 
The 
suitable 
and 
elegant 
space discretization of 
the NLS system (\ref{nonredNLS}) was 
proposed by 
Ablowitz and Ladik~\cite{AL1}
in the form
%
\begin{subequations}
\label{sd-AL}
\begin{align}
& \mathrm{i} q_{n,t} + (q_{n+1} + q_{n-1} -2 q_n)
        - q_n r_n (q_{n+1} + q_{n-1}) =0, \\
& \mathrm{i} r_{n,t} - (r_{n+1} + r_{n-1} -2 r_n)
        + r_n q_n (r_{n+1} + r_{n-1}) =0.
\end{align}
\end{subequations}
Indeed, 
system (\ref{sd-AL}) is 
integrable
and, 
with a rescaling of 
variables, 
reduces to (\ref{nonredNLS}) 
in 
the continuous
space limit. 
Moreover, (\ref{sd-AL}) 
allows the complex conjugation reduction 
between 
$q_n$ and $r_n$
and 
possesses 
the same invariance properties 
with respect to the space/time reflection 
as the continuous system (\ref{nonredNLS}). 
Although 
35 years have already passed since 
their work, 
the Ablowitz--Ladik
discretization 
(\ref{sd-AL}) 
is still 
a rare example 
of success. 
Indeed, 
even now, 
only a 
small number of suitable 
space discretizations of integrable 
NLS-type systems are known (see, {\it e.g.},~\cite{Tsuchi02}); 
they are all
\mbox{$(1+1)$}-dimensional 
systems with only one discrete spatial variable. 
The problem of 
how to discretize the continuous 
time variable in 
such 
%
systems 
is 
an 
interesting 
topic~\cite{2010JPA}, 
but we 
do 
not discuss it in this paper. 

The 
main objective of this paper is to provide 
the {\it first} example of a suitable 
discretization 
of an NLS-type system in \mbox{$2+1$} dimensions.
In particular, 
we 
consider 
the discretization 
of {\it both} spatial variables in 
a \mbox{$(2+1)$}-dimensional NLS system known as
the Davey--Stewartson system~\cite{DS74} 
(also see~\cite{Benney69}). 
Note that 
the Davey--Stewartson system
is integrable~\cite{Hab75,Morris77,Anker,Ab78,Cornille}
and 
appears to be 
the only genuinely \mbox{$(2+1)$}-dimensional 
generalization of the NLS system (\ref{nonredNLS})
(cf.~the Calogero--Degasperis system~\cite{Calo76}). 
Moreover, 
even if we 
include 
other types of 
integrable systems, 
the 
list of known 
systems with two discrete 
and one continuous 
independent variables 
is still 
very 
short. 
Thus, it is a highly nontrivial and 
challenging 
task 
to obtain 
the suitable 
space 
discretization
of the Davey--Stewartson system. 
%
%
To solve 
this problem, 
we first propose 
a 
natural 
\mbox{$(2+1)$}-dimensional generalization 
of 
the Ablowitz--Ladik lattice 
(cf.~(\ref{sd-AL})) 
by constructing its Lax pair. 
This \mbox{$(2+1)$}-dimensional Ablowitz--Ladik lattice 
certainly 
reduces to the Davey--Stewartson system 
in the continuous
space limit. 
A relevant Lax pair 
as well as 
the resulting system 
was previously 
%
studied 
by other authors~\cite{Hu06}
(also see~\cite{Hu07}), 
but the time part of our 
Lax pair
is essentially more general than 
the previously known one~\cite{Hu06}. 
As a result, 
the time evolution of 
our 
system 
is 
a linear 
combination 
of four elementary time evolutions, two of which 
were 
previously unknown. 
Moreover, it can be shown that the 
four time evolutions 
are mutually 
commutative.  
Thus, the \mbox{$(2+1)$}-dimensional Ablowitz--Ladik lattice 
is general enough and appears to be 
promising. 
However, it
does not allow the complex conjugation reduction 
directly 
and thus is not 
a suitable space 
discretization of 
the Davey--Stewartson system 
in 
its 
present form. 
To fix 
this shortcoming, 
we only have to 
consider 
a
certain nonlocal 
transformation of dependent variables, 
which 
symmetrizes the 
equations of motion. 
Thus, 
we 
obtain 
the suitable 
space discretization of the 
Davey--Stewartson system 
that indeed 
allows the complex conjugation reduction 
between the new 
variables 
after the transformation. 
In addition, 
the invariance properties of the continuous 
Davey--Stewartson system with respect to space/time reflections 
turn out to be properly
incorporated in
our 
space-discrete Davey--Stewartson system.

This paper is organized as follows. 
In section 2, we propose a \mbox{$(2+1)$}-dimensional version 
of the Ablowitz--Ladik lattice by considering an appropriate 
generalization of the 
Lax pair 
for the 
original Ablowitz--Ladik lattice.  
To uncover 
how the complex conjugation reduction 
can be imposed 
as an NLS-type system,  
we 
consider 
a nonlocal change 
of dependent variables; 
it can 
turn the \mbox{$(2+1)$}-dimensional Ablowitz--Ladik lattice 
into 
the suitable space discretization of 
the Davey--Stewartson system
that 
indeed 
allows the complex 
conjugation
reduction 
between the new 
dependent 
variables. 
In section 3, 
we elucidate how 
the general 
time evolution considered 
can be 
decomposed into 
four 
elementary 
time 
evolutions corresponding to the 
four 
directions 
on the two-dimensional lattice. 
On the basis of this decomposition 
and 
using the Hirota
bilinear method~\cite{Hirota04}, 
%
%
we 
construct 
some 
exact solutions 
of 
the \mbox{$(2+1)$}-dimensional Ablowitz--Ladik lattice 
and the space-discrete Davey--Stewartson system. 
In particular, 
multidromion solutions 
are presented explicitly. 
The last section, section 4, 
is devoted to concluding remarks.
%
%
%
%
%
%
%

\section{Derivation based on Lax pairs}

\subsection{\mbox{$(2+1)$}-dimensional Ablowitz--Ladik lattice}
\label{2.1}

As a generalization of the 
Lax pair
introduced 
by 
Ablowitz and Ladik~\cite{AL1}, 
we consider 
the following 
linear system 
on 
the 
two-dimensional lattice: 
\begin{subequations}
\label{DS1-Lax}
\begin{align}
& \boldsymbol{\Delta}_n^+ \psi_{n,m} := 
\psi_{n+1,m} -\psi_{n,m} = q_{n,m} \phi_{n,m},
\label{DS1-Lax1}
\\
& \boldsymbol{\Delta}_m^+ \phi_{n,m} :=
\phi_{n,m+1} -\phi_{n,m} = r_{n,m} \psi_{n,m},
\label{DS1-Lax2}
\\
& \frac{\partial \psi_{n,m}}{\partial t} 
  =a \psi_{n,m+1} + A_{n,m} \psi_{n,m} + C_{n,m} \psi_{n,m-1}
   + b q_{n-1,m} \phi_{n,m} - q_{n,m} D_{n,m} \phi_{n-1,m},
\label{DS1-Lax3}
\\
& \frac{\partial \phi_{n,m}}{\partial t} 
  =b \phi_{n+1,m} + B_{n,m} \phi_{n,m} + D_{n,m} \phi_{n-1,m}
   + a r_{n,m-1} \psi_{n,m} - r_{n,m} C_{n,m} \psi_{n,m-1}.
\label{DS1-Lax4}
\end{align}
\end{subequations}
Here, $\boldsymbol{\Delta}_n^+$ and $\boldsymbol{\Delta}_m^+$ 
denote the forward difference operators in 
each 
spatial 
direction, and 
the parameters 
$a$ and $b$ are arbitrary constants. 
The time dependence of the functions 
is 
usually 
suppressed. 
The compatibility conditions 
\mbox{$\partial_t \boldsymbol{\Delta}_n^+ \psi_{n,m}=
	\boldsymbol{\Delta}_n^+ \partial_t \psi_{n,m}$}
and \mbox{$ \partial_t \boldsymbol{\Delta}_m^+ \phi_{n,m}=
	\boldsymbol{\Delta}_m^+ \partial_t \phi_{n,m}$} 
for the linear system (\ref{DS1-Lax}) 
provide 
the 
\mbox{$(2+1)$}-dimensional Ablowitz--Ladik lattice, 
\begin{subequations}
\label{DS1}
\begin{align}
 \frac{\partial q_{n,m}}{\partial t} 
 &+ q_{n+1,m} D_{n+1,m} + b q_{n-1,m} - a q_{n,m+1} 
  - C_{n+1,m} q_{n,m-1}
\nonumber \\[-0.5mm] &
 + q_{n,m} B_{n,m} - A_{n+1,m} q_{n,m}=0,
\label{DS1-q}
\\[1mm]
 \frac{\partial r_{n,m}}{\partial t} 
 &+ r_{n,m+1} C_{n,m+1} + a r_{n,m-1} - b r_{n+1,m} - D_{n,m+1} r_{n-1,m}
\nonumber \\[-0.5mm] &
 + r_{n,m} A_{n,m} - B_{n,m+1} r_{n,m}=0,
\label{DS1-r}
\\[1mm]
& \hspace{-10mm}
A_{n+1,m} -A_{n,m} = -a (q_{n,m+1} r_{n,m} -q_{n,m} r_{n,m-1}),
\label{fdAL1}
\\[1mm]
& \hspace{-10mm}
B_{n,m+1} -B_{n,m} = -b (r_{n+1,m} q_{n,m} -r_{n,m} q_{n-1,m}), 
\label{fdAL2}
\\[1mm]
& \hspace{-10mm}
(1- q_{n,m} r_{n,m}) C_{n,m} =C_{n+1,m}(1- q_{n,m-1} r_{n,m-1}),
\label{fdAL3}
\\[1mm]
& \hspace{-10mm}
(1- r_{n,m} q_{n,m}) D_{n,m} =D_{n,m+1}(1- r_{n-1,m} q_{n-1,m}).
\label{fdAL4}
\end{align}
\end{subequations}
Indeed, 
if
all the functions 
depend on $n$ and $m$ only through 
\mbox{$n+m$}, 
(\ref{DS1}) reduces to 
the original 
Ablowitz--Ladik lattice~\cite{AL1}; 
the latter contains 
the integrable discrete NLS system (\ref{sd-AL}) as a special case. 
To restore 
the \mbox{$(1+1)$}-dimensional 
Lax pair 
involving
the spectral parameter $z$, we 
set 
\mbox{$\psi_{n,m}= z^{m-n} \psi'_{n,m}$} 
and \mbox{$\phi_{n,m}= z^{m-n-1} \phi'_{n,m}$}, 
rewrite the linear problem (\ref{DS1-Lax}) 
in terms of \mbox{$\psi'_{n,m}$} and \mbox{$\phi'_{n,m}$}, 
and then consider 
the dimensional reduction. 

Under appropriate boundary conditions at spatial infinity, 
we can 
use (\ref{fdAL1})--(\ref{fdAL4}) recursively 
to express the auxiliary 
fields 
$A_{n,m}$, $B_{n,m}$, $C_{n,m}$, and $D_{n,m}$ 
globally in terms of $q_{n,m}$ and $r_{n,m}$. 
Thus, they can be considered as the defining relations 
for the auxiliary fields, 
and 
the \mbox{$(2+1)$}-dimensional Ablowitz--Ladik lattice 
(\ref{DS1}) has intrinsically nonlocal nonlinearity. 
Note that 
relations
(\ref{fdAL1})--(\ref{fdAL4}) 
already appeared in 
the literature on
an integrable time discretization 
of the 
\mbox{$(1+1)$}-dimensional 
Ablowitz--Ladik lattice~\cite{AL77,Chiu77,Suris03}. 
Incidentally, in the stationary case of 
\mbox{$\partial_t q_{n,m} = \partial_t r_{n,m}=0$}, 
(\ref{DS1}) 
reduces to a nontrivial system of partial difference equations 
in \mbox{$1+1$} dimensions. 

Using the simple transformation
\[
q_{n,m} = q'_{n,m} \mathrm{e}^{\gamma t},\hspace{5mm}
r_{n,m} = r'_{n,m} \mathrm{e}^{-\gamma t},
\]
where $\gamma$ is a constant, and 
omitting the prime, 
we can introduce the terms $+\gamma q_{n,m}$ and $-\gamma r_{n,m}$
in (\ref{DS1-q}) and (\ref{DS1-r}), respectively. 
Clearly, 
these terms can be absorbed by 
constant shifts of $A_{n,m}$ and $B_{n,m}$. 

\subsection{Continuum limit}

By choosing the parameters appropriately 
and taking 
the continuous space limit, 
we can reduce system (\ref{DS1}) 
to the continuous 
Davey--Stewartson system. 
To see this, 
we first 
shift
the auxiliary fields 
as 
\begin{align}
& A_{n,m} = b-a + \widehat{A}_{n,m},\hspace{5mm}
  B_{n,m} = a-b + \widehat{B}_{n,m},
\nonumber \\
& C_{n,m} = a + \widehat{C}_{n,m},\hspace{5mm}
  D_{n,m} = b + \widehat{D}_{n,m},
\nonumber
\end{align}
and 
rewrite (\ref{DS1}) as 
\begin{subequations}
\label{DS-limit}
\begin{align}
\frac{\partial q_{n,m}}{\partial t} 
& + b (q_{n+1,m} + q_{n-1,m}-2q_{n,m})
 - a (q_{n,m+1} +q_{n,m-1}-2q_{n,m})  
\nonumber \\ & 
 + q_{n+1,m} \widehat{D}_{n+1,m} 
 - \widehat{C}_{n+1,m} q_{n,m-1}
 + q_{n,m} \widehat{B}_{n,m} - \widehat{A}_{n+1,m} q_{n,m}=0,
\label{}
\\
 \frac{\partial r_{n,m}}{\partial t} 
& + a (r_{n,m+1} +r_{n,m-1} -2r_{n,m})
  - b (r_{n+1,m}+ r_{n-1,m}-2r_{n,m})
 \nonumber \\ & 
  + r_{n,m+1} \widehat{C}_{n,m+1} - \widehat{D}_{n,m+1} r_{n-1,m}
  + r_{n,m} \widehat{A}_{n,m} - \widehat{B}_{n,m+1} r_{n,m}=0,
\label{}
\\
& \hspace{-10mm}
	\widehat{A}_{n+1,m} -\widehat{A}_{n,m} 
	= -a (q_{n,m+1} r_{n,m} -q_{n,m} r_{n,m-1}),
\label{}
\\
& \hspace{-10mm}
	\widehat{B}_{n,m+1} -\widehat{B}_{n,m} 
	= -b (r_{n+1,m} q_{n,m} -r_{n,m} q_{n-1,m}), 
\label{}
\\
& \hspace{-10mm}
 \widehat{C}_{n+1,m} - \widehat{C}_{n,m} 
 = - a(q_{n,m} r_{n,m} - q_{n,m-1} r_{n,m-1}) 
 +\widehat{C}_{n+1,m} q_{n,m-1} r_{n,m-1}- q_{n,m} r_{n,m} \widehat{C}_{n,m},
\\
& \hspace{-10mm}
 \widehat{D}_{n,m+1} - \widehat{D}_{n,m} 
 = - b (r_{n,m} q_{n,m}-r_{n-1,m} q_{n-1,m})
 + \widehat{D}_{n,m+1} r_{n-1,m} q_{n-1,m} - r_{n,m} q_{n,m}\widehat{D}_{n,m}.
\label{}
\end{align}
\end{subequations}
Subsequently, we rescale
the 
variables 
and 
parameters as 
\begin{align}
& q_{n,m}=\varDelta x \cdot q (x, y), \hspace{5mm}
 r_{n,m}=\varDelta y \cdot r (x,y), \hspace{5mm}
 x:=n \, \varDelta x , \hspace{5mm} y:=m\,\varDelta y,
\nonumber \\
& \widehat{A}_{n,m}= A (x,y), \hspace{5mm}
  \widehat{B}_{n,m}= B (x,y), \hspace{5mm}
  \widehat{C}_{n,m}= C (x,y), \hspace{5mm}
  \widehat{D}_{n,m}= D (x,y), 
\nonumber 
\end{align}
wherein the time dependence is suppressed
and 
\begin{align}
a =
\frac{\alpha}{(\varDelta y)^2}, \hspace{5mm}
b= \frac{\beta}{(\varDelta x)^2}.
\nonumber 
\end{align}
Thus, in the continuum limit 
\mbox{$\varDelta x, \varDelta y \to 0$}, 
(\ref{DS-limit}) reduces to the continuous 
Davey--Stewartson system~\cite{Ab78,Morris77,Cornille,Nizh80,Nizh82},
\begin{subequations}
\label{DS-conti}
\begin{align}
& q_t + \beta q_{xx}  
 - \alpha q_{yy}- (A+C) q + q (B+D) =0,
\label{}
\\
& r_t + \alpha r_{yy} - \beta r_{xx} + r (A+C) - (B+D) r =0,
\label{}
\\
& A_{x} =C_x = -\alpha (q r)_y,
\label{}
\\
& B_y = D_y = -\beta (rq)_x.
\label{}
\end{align}
\end{subequations}
Note that the Davey--Stewartson system (\ref{DS-conti}) 
is a linear combination
of the 
two commuting flows
corresponding to \mbox{$\alpha=0$}, \mbox{$\beta \neq0$} 
and \mbox{$\alpha \neq 0$}, \mbox{$\beta=0$}~\cite{Kaji90,MY97}
(also see~\cite{Nizh82}). 
In subsection~\ref{subs_commut}, 
we present 
its discrete analogue, 
that is, 
the \mbox{$(2+1)$}-dimensional Ablowitz--Ladik lattice (\ref{DS1})
is a linear combination of four
commuting flows. 
This 
is 
a 
quite 
natural 
result because 
(i) each of the two 
Davey--Stewartson flows 
provides 
an asymmetric 
\mbox{$(2+1)$}-dimensional generalization of 
the
NLS system and
(ii) the 
Ablowitz--Ladik 
discretization 
of the NLS system 
is actually 
a sum 
of two 
elementary 
flows 
(and 
one 
trivial 
flow) 
in the same 
hierarchy~\cite{AL1,Suris03,Kako,Kulish,GIK84}. 

\subsection{Noncommutative 
extension}

Actually, 
the \mbox{$(2+1)$}-dimensional Ablowitz--Ladik lattice 
(\ref{DS1}) is integrable 
in 
the general case 
where the dependent variables take their values in matrices, 
as long as 
the operations such as 
addition and 
multiplication
make sense. 
In that case, 
``$1$'' in (\ref{fdAL3}) 
and (\ref{fdAL4}) should be interpreted 
as the identity matrix. 

We can further 
generalize 
it to a variable-coefficient 
system 
wherein 
the parameters $a$ and $b$ become 
arbitrary matrix-valued 
functions of 
one 
spatial variable
as 
\mbox{$a_m:=a(m)$} and \mbox{$b_n:=b(n)$}. 
To obtain 
such an 
extension, 
we 
consider the 
following 
generalization of 
the 
linear system (\ref{DS1-Lax}):
\begin{subequations}
\label{nonDS1-Lax}
\begin{align}
\psi_{n+1,m} &= z \psi_{n,m} + q_{n,m} \phi_{n,m},
\label{nonDS1-Lax1}
\\[1mm]
\phi_{n,m+1} &= z^{-1} \phi_{n,m} + r_{n,m} \psi_{n,m},
\label{nonDS1-Lax2}
\\[1mm]
 \frac{\partial \psi_{n,m}}{\partial t} 
  &= z a_m \psi_{n,m+1} + A_{n,m} \psi_{n,m} 
  +z^{-1} C_{n,m} \psi_{n,m-1}
\nonumber \\[-0.5mm] & \hphantom{=} \;
  +z^{-1} q_{n-1,m} b_{n-1} \phi_{n,m} - q_{n,m} D_{n,m} \phi_{n-1,m},
\label{nonDS1-Lax3}
\\[1mm]
 \frac{\partial \phi_{n,m}}{\partial t} 
  & =z^{-1} b_n \phi_{n+1,m} + B_{n,m} \phi_{n,m} 
	+ z D_{n,m} \phi_{n-1,m}
\nonumber \\[-0.5mm] & \hphantom{=} \;
   + z r_{n,m-1} a_{m-1} \psi_{n,m} - r_{n,m} C_{n,m} \psi_{n,m-1}.
\label{nonDS1-Lax4}
\end{align}
\end{subequations}
%
Note that the
``spectral parameter'' 
$z$ is 
nonessential in the \mbox{$(2+1)$}-dimensional case 
and can be 
fixed at $1$ 
as described in subsection~\ref{2.1}. 
The compatibility conditions for
the linear system (\ref{nonDS1-Lax}) 
indeed provide 
the 
noncommutative 
system with 
site-dependent 
coefficients, 
%
\begin{subequations}
\label{nonDS1}
\begin{align}
\frac{\partial q_{n,m}}{\partial t} 
 & + q_{n+1,m} D_{n+1,m} + q_{n-1,m} b_{n-1} - a_m q_{n,m+1} 
  - C_{n+1,m} q_{n,m-1}
\nonumber \\ & 
 + q_{n,m} B_{n,m} - A_{n+1,m} q_{n,m}=0,
\label{}
\\[1mm]
\frac{\partial r_{n,m}}{\partial t} 
& +r_{n,m+1} C_{n,m+1} +r_{n,m-1}a_{m-1} - b_n r_{n+1,m} - D_{n,m+1} r_{n-1,m}
\nonumber \\ & 
 + r_{n,m} A_{n,m} - B_{n,m+1} r_{n,m}=0,
\label{}
\\[1mm]
& \hspace{-10mm}
A_{n+1,m} -A_{n,m} = -a_m q_{n,m+1} r_{n,m} + q_{n,m} r_{n,m-1}a_{m-1},
\label{}
\\[1mm]
& \hspace{-10mm}
B_{n,m+1} -B_{n,m} = -b_n r_{n+1,m} q_{n,m} +r_{n,m} q_{n-1,m}b_{n-1}, 
\label{}
\\[1mm]
& \hspace{-10mm}
(I- q_{n,m} r_{n,m}) C_{n,m} =C_{n+1,m}(I- q_{n,m-1} r_{n,m-1}),
\label{I-e}
\\[1mm]
& \hspace{-10mm}
(I- r_{n,m} q_{n,m}) D_{n,m} =D_{n,m+1}(I- r_{n-1,m} q_{n-1,m}).
\label{I-f}
\end{align}
\end{subequations}
If $q_{n,m}$ and $r_{n,m}$ are rectangular 
matrices, 
the identity matrix $I$ in (\ref{I-e}) and that 
in (\ref{I-f}) 
have unequal
sizes. 
In the commutative case of the parameters, 
the site-dependent 
nature of (\ref{nonDS1})
is nonessential if both 
\mbox{$\prod_{m=-\infty}^\infty a_m$} and 
\mbox{$\prod_{n=-\infty}^\infty b_n$} take nonzero 
finite values. 
Indeed, if 
we change the variables as 
\begin{align}
& q_{n,m}= 
\left( \prod_{j=-\infty}^{m-1} a_j \right)^{-1}
 \left( \prod_{k=-\infty}^{n-1} b_k\right)
	\widetilde{q}_{n,m}, \hspace{5mm}
r_{n,m}= 
\left( \prod_{j=-\infty}^{m-1} a_j \right)
 \left( \prod_{k=-\infty}^{n-1} b_k\right)^{-1}
	\widetilde{r}_{n,m}, \hspace{5mm}
\nonumber \\[1mm]
& C_{n,m} = a_{m-1}^{-1}\widetilde{C}_{n,m}, 
\hspace{5mm} D_{n,m} = b_{n-1}^{-1}\widetilde{D}_{n,m},
\nonumber 
\end{align}
the site-dependent 
parameters 
can be normalized to 
$1$. 
We can 
also 
obtain a similar result 
in the noncommutative case.

\subsection{Appropriate 
change of dependent 
variables}
\label{2.4}

For simplicity, in the following discussion, 
we consider only the 
commutative and constant-coefficient
case 
wherein the parameters $a$ and $b$ are constants 
and all the quantities are scalar. 
Thus, 
the lowest-order conservation law for (\ref{DS1}) is given by 
\begin{align}
 \frac{\partial \log ( 1- q_{n,m}r_{n,m} )}{\partial t} 
&= \boldsymbol{\Delta}_n^+ 
\left[ -b q_{n-1,m}r_{n,m} 
 + D_{n,m}(1- q_{n-1,m}r_{n-1,m} )^{-1}q_{n,m} r_{n-1,m} \right]
\nonumber \\ & 
+ \boldsymbol{\Delta}_m^+
\left[ - a q_{n,m} r_{n,m-1} 
 +  C_{n,m}(1- q_{n,m-1} r_{n,m-1})^{-1}q_{n,m-1}r_{n,m} \right].
\label{DScons}
\end{align}
The existence of an ultralocal conserved density 
\mbox{$\log ( 1- q_{n,m}r_{n,m} )$} 
implies that 
a 
nonlocal 
transformation involving infinite products 
of \mbox{$(1- q_{n,m}r_{n,m})^\delta$} 
with \mbox{$\delta \neq 0$} 
could be 
applied (cf.~\cite{GI2}); 
this is indeed the case as we 
will 
see below. 

The \mbox{$(2+1)$}-dimensional Ablowitz--Ladik lattice (\ref{DS1}) 
is invariant under a
space reflection 
\mbox{$(n,m) 
\to (-m,-n)$} 
with a minor redefinition 
of the parameters and 
the auxiliary fields. 
However, (\ref{DS1}) 
does not 
allow 
the complex conjugation reduction 
between 
$q_{n,m}$ and $r_{n,m}$
in the local form. 
Therefore, 
we need to 
identify 
new ``conjugate'' variables 
instead of $q_{n,m}$ and $r_{n,m}$ and 
rewrite 
(\ref{DS1}) in 
a more symmetric form 
using 
the new 
variables. 
For this purpose, 
we 
consider a gauge transformation 
so that the spatial part of the Lax representation 
obtains 
invariance 
with respect to the combined space reflection 
\mbox{$(n,m) \to (-n,-m)$} 
or, equivalently, a
180 degree 
rotation 
around the origin. 
Thus, we apply 
the gauge transformation 
\begin{equation}
\psi_{n,m} = X_{n,m}
\Psi_{n,m}, 
\hspace{5mm}
\phi_{n,m} = Y_{n,m}
\Phi_{n,m}
\label{gauge1}
\end{equation}
to (\ref{DS1-Lax1}) and (\ref{DS1-Lax2}) 
and 
change the dependent variables as 
\begin{equation}
u_{n,m} = \frac{Y_{n,m}
}{X_{n,m}
}q_{n,m}, \hspace{5mm}
v_{n,m} = \frac{X_{n,m}
}{Y_{n,m}
} r_{n,m}.
\label{trans1}
\end{equation}
Here, $X_{n,m}$ and $Y_{n,m}$ are defined as 
\begin{align}
\label{XY}
& X_{n,m}:=\frac{1}{h_m} \prod_{j=-\infty}^{n-1} \sqrt{1- q_{j,m} r_{j,m}},
\hspace{5mm}
Y_{n,m}:= \frac{1}{l_n} \prod_{k=-\infty}^{m-1} \sqrt{1- q_{n,k} r_{n,k}}.
\end{align}
The 
norming 
functions 
$h_m (t)$ and $l_n (t)$ are introduced to 
realize the complex conjugation reduction 
between $u_{n,m}$ and $v_{n,m}$; they 
will be 
determined 
later.
%
One can also use 
\mbox{$\left(
\prod_{k=m}^{+\infty} \sqrt{1- q_{n,k} r_{n,k}} \right)^{-1}$}
instead of \mbox{$\prod_{k=-\infty}^{m-1} \sqrt{1- q_{n,k} r_{n,k}}$} 
to maintain the invariance 
under the space reflection 
\mbox{$(n,m) 
\to (-m,-n)$}. 
This modification
causes no essential difference in the following discussion, 
so 
the transformed system 
can become 
ideally symmetric 
with respect to 
space reflections. 
Here and hereafter, 
we assume that 
\mbox{$\abs{q_{n,m} r_{n,m}} \ll 1$} 
so that \mbox{$\sqrt{1- q_{n,m} r_{n,m}}$} 
and its inverse as well as their infinite products as considered 
above 
are 
uniquely 
and 
well defined. 
For example, we 
consider that 
\mbox{$q_{n,m} 
=
O (\varDelta x)$} and \mbox{$r_{n,m} = O (\varDelta y)$} 
(cf.~(\ref{DS1-Lax1}) and (\ref{DS1-Lax2})), 
and 
\mbox{$\sqrt{1- q_{n,m} r_{n,m}}$} 
is defined 
as 
the Maclaurin series
in \mbox{$q_{n,m} r_{n,m} = O(\varDelta x \hspace{1pt}\varDelta y)$}.
Note that \mbox{$u_{n,m} v_{n,m}=q_{n,m} r_{n,m}$}, so that 
the inverse transformation of (\ref{trans1}) can be obtained immediately. 
Thus, the spatial part of the Lax representation 
acquires the form 
\begin{equation}
\left[
\begin{array}{c}
 \Psi_{n+1,m} \\
 \Phi_{n,m+1} \\
\end{array}
\right]
= \frac{1}{\sqrt{1- u_{n,m} v_{n,m}}}
\left[
\begin{array}{cc}
1 &  u_{n,m} \\
v_{n,m} & 1 \\
\end{array}
\right]
\left[
\begin{array}{c}
 \Psi_{n,m} \\
 \Phi_{n,m} \\
\end{array}
\right].
\label{DS2-Lax1}
\end{equation}
Very recently, 
D.~Zakharov has 
considered essentially the same scattering 
problem
in~\cite{Dmitry09-1,Dmitry09-2}. 
However, 
this is an accidental coincidence, because 
the first author
arrived at 
this Lax representation as well as its generalization 
implied 
in 
subsection~\ref{2.5}
independently
before 
the papers~\cite{Dmitry09-1,Dmitry09-2} appeared. 
The invariance 
under the combined space reflection
\mbox{$(n,m) \to (-n,-m)$}
can be 
easily 
seen if we shift 
the indices of the linear wavefunction by $1/2$, i.e., 
\[
\left[
\begin{array}{c}
 \widehat{\Psi}_{n+\frac{1}{2},m} \\
 \widehat{\Phi}_{n,m+\frac{1}{2}} \\
\end{array}
\right]
= \frac{1}{\sqrt{1- u_{n,m} v_{n,m}}}
\left[
\begin{array}{cc}
1 &  u_{n,m} \\
v_{n,m} & 1 \\
\end{array}
\right]
\left[
\begin{array}{c}
 \widehat{\Psi}_{n-\frac{1}{2},m} \\
 \widehat{\Phi}_{n,m-\frac{1}{2}} \\
\end{array}
\right],
\]
where \mbox{$\Psi_{n,m}=:\widehat{\Psi}_{n-\frac{1}{2},m}$} 
and \mbox{$\Phi_{n,m} =:\widehat{\Phi}_{n,m-\frac{1}{2}}$}. 
Indeed, because the determinant of 
the spatial Lax matrix above 
is unity,
we obtain
\[
\left[
\begin{array}{c}
 \widehat{\Psi}_{n-\frac{1}{2},m} \\
 - \widehat{\Phi}_{n,m-\frac{1}{2}} \\
\end{array}
\right]
= \frac{1}{\sqrt{1- u_{n,m} v_{n,m}}}
\left[
\begin{array}{cc}
1 &  u_{n,m} \\
v_{n,m} & 1 \\
\end{array}
\right]
\left[
\begin{array}{c}
 \widehat{\Psi}_{n+\frac{1}{2},m} \\
 -\widehat{\Phi}_{n,m+\frac{1}{2}} \\
\end{array}
\right].
\]

It should be possible to apply 
the transformation 
(\ref{trans1}) with (\ref{XY})
directly 
to 
the \mbox{$(2+1)$}-dimensional Ablowitz--Ladik lattice 
(\ref{DS1}) 
and 
derive the transformed 
equations of motion 
with the aid of the conservation law (\ref{DScons}). 
However, 
the 
nonlocal nature of (\ref{DS1})
makes such a computation 
rather complicated and difficult. 
%
Thus, 
as an 
alternative, 
we 
apply 
the gauge transformation (\ref{gauge1}) 
with (\ref{trans1}) and (\ref{XY})
to the time 
part of the Lax representation, 
(\ref{DS1-Lax3}) and (\ref{DS1-Lax4}), and determine 
the time evolution of 
the 
gauge-transformed 
wavefunction, 
$\partial_t \Psi_{n,m}$ and $\partial_t \Phi_{n,m}$. 
Its compatibility 
with 
the scattering problem 
(\ref{DS2-Lax1})
can 
provide
the transformed 
equations of motion. 

\subsection{Space-discrete 
Davey--Stewartson system}
\label{2.5}

Before applying 
the 
transformation described 
in subsection~\ref{2.4}, 
we 
fix 
the 
boundary conditions for 
the \mbox{$(2+1)$}-dimensional Ablowitz--Ladik lattice 
(\ref{DS1}) 
as 
\begin{subequations}
\label{ALbc1}
\begin{align}
& \lim_{n \to - \infty} (q_{n,m}, r_{n,m}) = 
\lim_{m \to - \infty} (q_{n,m}, r_{n,m}) = \vt{0}, 
\label{qr-bc}
\\[1mm]
& \lim_{n \to - \infty} C_{n,m} = c \left( \frac{h_{m-1}}{h_{m}}\right)^2, 
\hspace{5mm}
\lim_{m \to - \infty} D_{n,m} = d \left( \frac{l_{n-1}}{l_{n}}\right)^2. 
\label{ALbc1-CD}
\end{align}
\end{subequations}
However, 
we do not fix 
\mbox{$\lim_{n \to - \infty} A_{n,m}$} 
and 
\mbox{$\lim_{m \to - \infty} B_{n,m}$} 
in order 
to obtain interesting solutions such as dromion solutions; 
they 
can 
also depend on 
the remaining spatial variable and 
time $t$. 
In (\ref{qr-bc}), the dynamical variables 
$q_{n,m}$ and $r_{n,m}$ are assumed to approach zero 
sufficiently rapidly. 
In (\ref{ALbc1-CD}), 
$c$ and $d$ are constants. 
The defining 
relations 
(\ref{fdAL3}) and (\ref{fdAL4}) 
enable $C_{n,m}$ and $D_{n,m}$ to be expressed globally 
as 
\[
C_{n,m}= c \left(\frac{X_{n,m}}{X_{n,m-1}}\right)^2, 
\hspace{5mm}
D_{n,m}= d \left(\frac{Y_{n,m}}{Y_{n-1,m}}\right)^2.
\]
Thus, the gauge transformation (\ref{gauge1})
with (\ref{trans1}) and (\ref{XY})
changes (\ref{DS1-Lax3}) and (\ref{DS1-Lax4}) to
\begin{align}
 \frac{\partial \Psi_{n,m}}{\partial t}
 &=  a\frac{X_{n,m+1}}{X_{n,m}} \Psi_{n,m+1}
 + \widetilde{A}_{n,m} \Psi_{n,m} 
 + c \frac{X_{n,m}}{X_{n,m-1}} \Psi_{n,m-1}
\nonumber \\[1mm] & {\hphantom =}\;
 + b \frac{Y_{n,m}}{Y_{n-1,m+1}} u_{n-1,m}\Phi_{n,m} 
-d \frac{Y_{n,m}}{Y_{n-1,m}} u_{n,m}\Phi_{n-1,m}
\nonumber
\end{align}
and
\begin{align}
 \frac{\partial \Phi_{n,m}}{\partial t}
 &=b\frac{Y_{n+1,m}}{Y_{n,m}} \Phi_{n+1,m} +\widetilde{B}_{n,m} \Phi_{n,m} 
 +d\frac{Y_{n,m}}{Y_{n-1,m}}\Phi_{n-1,m}
\nonumber \\[1mm] & {\hphantom =}\;
 +a \frac{X_{n,m}}{X_{n+1,m-1}}v_{n,m-1} \Psi_{n,m} 
- c \frac{X_{n,m}}{X_{n,m-1}} v_{n,m}\Psi_{n,m-1},
\nonumber
\end{align}
where 
\begin{align}
\label{trans2}
& \widetilde{A}_{n,m} := A_{n,m}
	 - \frac{\partial_t X_{n,m}}{X_{n,m}},\hspace{5mm}
\widetilde{B}_{n,m} := B_{n,m} - \frac{\partial_t Y_{n,m}}{Y_{n,m}}.
\end{align}
Recalling that \mbox{$q_{n,m} r_{n,m}=u_{n,m} v_{n,m}$}, 
the above relations 
combined with (\ref{DS2-Lax1}) 
comprise 
the Lax representation for the transformed system. 
To express it 
in a concise form,
we introduce 
the quantities 
\begin{subequations}
\label{trans3}
\begin{align}
& w_{n,m}:= \frac{1}{\sqrt{1- q_{n,m} r_{n,m}}}
	= \frac{1}{\sqrt{1- u_{n,m} v_{n,m}}},
\label{} \\[2mm]
& f_{n,m}:= \frac{X_{n,m+1}}{X_{n,m}}
 = \frac{h_{m} \prod_{j=-\infty}^{n-1} \sqrt{1- u_{j,m+1} v_{j,m+1}}}
        {h_{m+1} \prod_{j=-\infty}^{n-1} \sqrt{1- u_{j,m} v_{j,m}}}, 
\label{} \\[2mm] & 
g_{n,m}:= \frac{Y_{n+1,m}}{Y_{n,m}}
 = \frac{l_n \prod_{k=-\infty}^{m-1} \sqrt{1- u_{n+1,k} v_{n+1,k}}}
 {l_{n+1} \prod_{k=-\infty}^{m-1} \sqrt{1- u_{n,k} v_{n,k}}}.
\end{align}
\end{subequations}
Thus, 
we obtain 
the Lax representation in the form,
\begin{subequations}
\label{dDS-Lax}
\begin{align}
 \Psi_{n+1,m} &= w_{n,m} \Psi_{n,m} + w_{n,m} u_{n,m} \Phi_{n,m},
\label{dDS-Lax1}
\\[1mm]
 \Phi_{n,m+1} &= w_{n,m} \Phi_{n,m} + w_{n,m} v_{n,m} \Psi_{n,m},
\label{dDS-Lax2}
\\[1mm]
 \frac{\partial \Psi_{n,m}}{\partial t}
 &= a f_{n,m} \Psi_{n,m+1} + \widetilde{A}_{n,m} \Psi_{n,m} 
	+c f_{n,m-1} \Psi_{n,m-1}
\nonumber \\[-1mm] & \hphantom{=}\;\hspace{1pt}
 + b w_{n-1,m} g_{n-1,m} u_{n-1,m}\Phi_{n,m}- d g_{n-1,m}u_{n,m}\Phi_{n-1,m},
\label{dDS-Lax3}
\\[1mm]
 \frac{\partial \Phi_{n,m}}{\partial t}
 & =b g_{n,m} \Phi_{n+1,m} + \widetilde{B}_{n,m} \Phi_{n,m} 
	+ d g_{n-1,m} \Phi_{n-1,m}
\nonumber \\[-1mm] & \hphantom{=}\;\hspace{1pt}
 +a w_{n,m-1}f_{n,m-1}v_{n,m-1} \Psi_{n,m} - c f_{n,m-1} v_{n,m} \Psi_{n,m-1}.
\label{dDS-Lax4}
\end{align}
\end{subequations}
The corresponding 
boundary conditions are given by 
\begin{align}
& \lim_{n \to - \infty} (u_{n,m}, v_{n,m}) =
\lim_{m \to - \infty} (u_{n,m}, v_{n,m}) = \vt{0},
\nonumber
\\[1mm]
& \lim_{n \to - \infty} f_{n,m} = \frac{h_m}{h_{m+1}}, \hspace{5mm}
\lim_{m \to - \infty} g_{n,m} = \frac{l_n}{l_{n+1}}.
\nonumber
\end{align}

Actually, 
we can generalize 
(\ref{dDS-Lax}) to 
a more general 
form
wherein the spatial part 
is 
given by 
\begin{align}
& \Psi_{n+1,m} = w_{n,m} \Psi_{n,m} + q_{n,m} \Phi_{n,m},
\nonumber 
\\
& \Phi_{n,m+1} = s_{n,m} \Phi_{n,m} + r_{n,m} \Psi_{n,m},
\nonumber 
\end{align}
with four independent functions 
$w_{n,m}$, $s_{n,m}$, $q_{n,m}$, and $r_{n,m}$.
Thus, it is possible to start with this general 
Lax representation and then consider the 
reduction. 
However, we 
skip such a discussion 
to
maintain an
easy-to-read flow of the paper.

The compatibility conditions 
for the linear system (\ref{dDS-Lax}) with 
\mbox{$ w_{n,m}=\left( 1- u_{n,m} v_{n,m} \right)^{-\frac{1}{2}}$}
provide the 
time evolution equations 
for $u_{n,m}$ and $v_{n,m}$.
They can be 
written in a 
natural 
compact form 
using 
new auxiliary fields 
$\alpha_{n,m}$ and $\beta_{n,m}$ defined 
as 
\begin{subequations}
\label{trans4}
\begin{align}
& \widetilde{A}_{n,m} 
 =: \frac{1}{2} w_{n-1,m} g_{n-1,m}\left( b u_{n-1,m}v_{n,m}
	-d u_{n,m} v_{n-1,m} \right)-\frac{1}{2} \alpha_{n,m},
\label{} \\[1mm]
& \widetilde{B}_{n,m} 
 =: \frac{1}{2} w_{n,m-1} f_{n,m-1} \left( a u_{n,m}v_{n,m-1}
	-c u_{n,m-1} v_{n,m}\right) -\frac{1}{2} \beta_{n,m}. 
\label{}
\end{align}
\end{subequations}
Thus, 
%
we finally 
arrive at 
the desired system, 
\begin{subequations}
\label{sym-dDS}
\begin{align}
& \frac{\partial u_{n,m}}{\partial t}
 + (1-u_{n,m}v_{n,m}) \left( d w_{n,m} g_{n,m}u_{n+1,m} 
 + b w_{n-1,m} g_{n-1,m} u_{n-1,m} \right.
\nonumber \\ & 
\hspace{27mm}
\left. \mbox{}- a w_{n,m} f_{n,m} u_{n,m+1} - c w_{n,m-1}f_{n,m-1} u_{n,m-1}
\right)
\nonumber \\ & 
\hspace{10mm}
+\frac{1}{2} u_{n,m}\left[ \alpha_{n,m} -w_{n,m-1} f_{n,m-1} 
 (a u_{n,m} v_{n,m-1} +c u_{n,m-1} v_{n,m}) \right.
\nonumber \\ & 
\hspace{27mm} 
\left. \mbox{}- \beta_{n,m} + w_{n-1,m} g_{n-1,m} (b u_{n-1,m} v_{n,m} 
 +d u_{n,m}v_{n-1,m})\right]=0,
\label{symm-1}
\\[1mm]
& \frac{\partial v_{n,m}}{\partial t}
 + (1-u_{n,m}v_{n,m}) \left( c w_{n,m} f_{n,m} v_{n,m+1} 
 + a w_{n,m-1}f_{n,m-1} v_{n,m-1}
\right.
\nonumber \\ & 
\hspace{27mm}
\left. \mbox{} -b w_{n,m} g_{n,m}v_{n+1,m} 
 -d w_{n-1,m} g_{n-1,m} v_{n-1,m} 
\right)
\nonumber \\ & 
\hspace{10mm}
-\frac{1}{2} v_{n,m}\left[ \alpha_{n,m} -w_{n,m-1} f_{n,m-1} 
 (a u_{n,m} v_{n,m-1} +c u_{n,m-1} v_{n,m}) \right.
\nonumber \\ & 
\hspace{27mm} 
\left. \mbox{}- \beta_{n,m} + w_{n-1,m} g_{n-1,m} (b u_{n-1,m} v_{n,m} 
 +d u_{n,m}v_{n-1,m})\right]=0,
\label{symm-2}
\\[1mm]
& w_{n,m}f_{n,m}=w_{n,m+1}f_{n+1,m} \hspace{3mm} \mathrm{if} \; 
	(a,c) \neq \vt{0},
\label{symm-3}
\\[1mm]
& w_{n,m}g_{n,m}=w_{n+1,m}g_{n,m+1} \hspace{3mm} \mathrm{if} \; 
	(b,d) \neq \vt{0},
\label{symm-4}
\\[1mm]
& \boldsymbol{\Delta}_n^+ \alpha_{n,m}
= \boldsymbol{\Delta}_m^+ \left[ w_{n,m-1} f_{n,m-1}
 (a u_{n,m} v_{n,m-1} + c u_{n,m-1}v_{n,m})\right],
\label{symm-5} \\[1mm]
& \boldsymbol{\Delta}_m^+ \beta_{n,m}
= \boldsymbol{\Delta}_n^+ \left[ w_{n-1,m} g_{n-1,m}
 (b u_{n-1,m} v_{n,m} + d u_{n,m}v_{n-1,m})\right].
\label{symm-6} 
\end{align}
\end{subequations}
Here, $a$, $b$, $c$, and $d$ are constants and 
\mbox{$ w_{n,m}=\left( 1- u_{n,m} v_{n,m} \right)^{-\frac{1}{2}}$}. 
In the same way as (\ref{DS1}), 
(\ref{sym-dDS}) 
also admits
a dimensional reduction 
to the 
Ablowitz--Ladik lattice~\cite{AL1}. 
Using 
(\ref{symm-3})--(\ref{symm-6}),
we can
rewrite 
(\ref{symm-1}) and (\ref{symm-2}) 
in
a more symmetric 
form 
with respect to 
space reflections.
When \mbox{$c= -a^\ast$} and \mbox{$d= -b^\ast$}, 
the \mbox{$(2+1)$}-dimensional system (\ref{sym-dDS}) allows 
the complex conjugation reduction 
\mbox{$v_{n,m}=\sigma u_{n,m}^\ast$} with a real constant $\sigma$; 
in this reduction, 
the auxiliary fields $f_{n,m}$ and $g_{n,m}$ 
become 
real-valued, 
while the auxiliary fields $\alpha_{n,m}$ and $\beta_{n,m}$ 
become 
purely imaginary. 
In particular, 
(\ref{sym-dDS}) with 
purely imaginary $a$, $b$, 
\mbox{$c\hspace{1pt}(= -a^\ast)$}, and \mbox{$d\hspace{1pt}(= -b^\ast)$} 
provides the suitable space discretization 
of the Davey--Stewartson system (cf.~(\ref{DS-conti})). 

Similarly to the continuous case 
(cf.~\cite{Bogdanov1,Bogdanov2,Boiti88}), 
when 
\mbox{$c= -a$} and \mbox{$d= -b$}, 
we can consider the 
reduction of
\mbox{$v_{n,m}=\sigma u_{n,m}$}
and
\mbox{$\alpha_{n,m}=\beta_{n,m}=0$}
to obtain 
a \mbox{$(2+1)$}-dimensional analogue of 
the modified Volterra lattice~\cite{Hiro73}, 
\begin{subequations}
\label{2DmLV}
\begin{align}
& \frac{\partial u_{n,m}}{\partial t} = 
  \left( 1-\sigma u_{n,m}^2 \right) \left( b w_{n,m} g_{n,m}u_{n+1,m} 
 - b w_{n-1,m} g_{n-1,m} u_{n-1,m} \right.
\nonumber \\ & 
\hspace{27mm}
\left. \mbox{}+ a w_{n,m} f_{n,m} u_{n,m+1} -a w_{n,m-1}f_{n,m-1} u_{n,m-1}
\right),
\label{}
\\[1mm]
& w_{n,m}f_{n,m}=w_{n,m+1}f_{n+1,m} \hspace{3mm} \mathrm{if} \; 
	a \neq 0,
\label{}
\\[1mm]
& w_{n,m}g_{n,m}=w_{n+1,m}g_{n,m+1} \hspace{3mm} \mathrm{if} \; 
	b \neq 0.
\label{}
\end{align}
\end{subequations}
Here, \mbox{$ w_{n,m}=\left( 1- \sigma u_{n,m}^2 \right)^{-\frac{1}{2}}$}. 
It would be interesting to 
look 
for a relationship between 
(\ref{2DmLV}) and 
the discrete 
modified Nizhnik--Veselov--Novikov hierarchy  
in~\cite{Dmitry09-1} (also see~\cite{HuJPA05,HuJMAA05,Kri10}).

\section{Solutions 
by the Hirota 
method}

In this section, we 
discuss 
how to 
construct 
exact solutions of the 
discrete Davey--Stewartson system (\ref{sym-dDS}) 
using the Hirota bilinear method~\cite{Hirota04}. 
Because of 
the 
complexity 
and irrationality
of the equations of motion, 
it would be
too hard 
to 
solve (\ref{sym-dDS}) directly, 
so 
we take 
an alternative approach. 
First, 
we 
bilinearize 
the \mbox{$(2+1)$}-dimensional Ablowitz--Ladik lattice (\ref{DS1}). 
Subsequently, we 
consider 
the effect of 
the 
nonlocal 
transformation (\ref{trans1}) 
with (\ref{XY}), 
(\ref{trans2}), (\ref{trans3}), and (\ref{trans4}) 
in the bilinear formalism. 
The 
infinite products
appearing in the 
nonlocal transformation 
can essentially be 
expressed locally 
in terms of 
a ``tau function''.
Thus, 
we can 
obtain 
exact solutions of (\ref{DS1}) 
and (\ref{sym-dDS}) concurrently 
from the same set of bilinear equations. 

\subsection{Decomposition into 
four commutative
flows}
\label{subs_commut}

Before applying 
the Hirota bilinear method, 
we demonstrate that 
the \mbox{$(2+1)$}-dimensional Ablowitz--Ladik lattice (\ref{DS1}) 
can be decomposed into the four elementary flows. 
In view of (\ref{fdAL1}), 
(\ref{fdAL2}), and 
(\ref{ALbc1-CD}), 
we 
rescale the auxiliary fields 
as
\begin{equation}
\label{aux-scale}
A_{n,m}=: a A^{(0)}_{n,m}, \hspace{5mm}
B_{n,m}=: b B^{(0)}_{n,m}, \hspace{5mm}
C_{n,m}=: c \hspace{1pt}C^{(0)}_{n,m}, \hspace{5mm}
D_{n,m}=: d D^{(0)}_{n,m}. 
\end{equation}
The corresponding boundary conditions are 
\begin{align}
& \lim_{n \to - \infty} (q_{n,m}, r_{n,m}) =
\lim_{m \to - \infty} (q_{n,m}, r_{n,m}) = \vt{0},
\nonumber
\\[1mm]
& \lim_{n \to - \infty} C_{n,m}^{(0)} = 
\left( \frac{h_{m-1}}{h_{m}}\right)^2, \hspace{5mm}
\lim_{m \to - \infty} D_{n,m}^{(0)} = 
\left( \frac{l_{n-1}}{l_{n}}\right)^2.
\nonumber
\end{align}
Thus, 
considering the 
simplest 
cases where only one of the 
parameters $a$, $b$, $c$, and $d$ 
does not vanish, we 
obtain the four elementary systems: 
\\
\vspace{-3mm}
\\
$\bullet$ $a$-system
\begin{subequations}
\label{a-system}
\begin{align}
& \partial_{t_a} q_{n,m} =  q_{n,m+1} + A_{n+1,m}^{(0)} q_{n,m},
\label{}
\\
& \partial_{t_a} r_{n,m} =-  r_{n,m-1} - r_{n,m} A_{n,m}^{(0)},
\label{}
\\
& A^{(0)}_{n+1,m} -A^{(0)}_{n,m} = 
 -\left( q_{n,m+1} r_{n,m} -q_{n,m} r_{n,m-1} \right),
\label{}
\end{align}
\end{subequations}
$\bullet$ $b$-system
\begin{subequations}
\label{b-system}
\begin{align}
& \partial_{t_b} q_{n,m} =
  - q_{n-1,m} - q_{n,m} B_{n,m}^{(0)},
\label{}
\\
& \partial_{t_b} r_{n,m}
  = r_{n+1,m} + B_{n,m+1}^{(0)} r_{n,m},
\label{}
\\
& B_{n,m+1}^{(0)} -B_{n,m}^{(0)} 
 = -\left( r_{n+1,m} q_{n,m} -r_{n,m} q_{n-1,m} \right),
\label{}
\end{align}
\end{subequations}
$\bullet$ $c$-system
\begin{subequations}
\label{c-system}
\begin{align}
 & \partial_{t_c} q_{n,m} = C_{n+1,m}^{(0)} q_{n,m-1},
\label{}
\\
& \partial_{t_c} r_{n,m} =- r_{n,m+1} C_{n,m+1}^{(0)},
\label{}
\\
& (1- q_{n,m} r_{n,m}) C_{n,m}^{(0)} =C_{n+1,m}^{(0)}(1- q_{n,m-1} r_{n,m-1}),
\label{}
\end{align}
\end{subequations}
$\bullet$ $d$-system
\begin{subequations}
\label{d-system}
\begin{align}
 & \partial_{t_d} q_{n,m} =- q_{n+1,m} D_{n+1,m}^{(0)},
\label{}
\\
 & \partial_{t_d} r_{n,m} = D_{n,m+1}^{(0)} r_{n-1,m},
\label{}
\\
& (1- r_{n,m} q_{n,m}) D_{n,m}^{(0)}=D_{n,m+1}^{(0)}(1- r_{n-1,m} q_{n-1,m}).
\label{}
\end{align}
\end{subequations}
%
Clearly, 
the time evolution in 
(\ref{DS1}) is a linear combination of these four 
time evolutions, that is, 
\mbox{$\partial_t = a\partial_{t_a}+b\partial_{t_b}
	+c\partial_{t_c}+d\partial_{t_d}$}. 
In fact, they 
are 
mutually 
commutative, 
so 
the above 
four systems 
belong to the same integrable hierarchy
as the original system (\ref{DS1}). 
To check 
the commutativity conditions 
\mbox{$\partial_{t_\alpha} \partial_{t_\beta} q_{n,m}
= \partial_{t_\beta} \partial_{t_\alpha} q_{n,m}$} and
\mbox{$\partial_{t_\alpha} \partial_{t_\beta} r_{n,m}
= \partial_{t_\beta} \partial_{t_\alpha} r_{n,m}$}
for \mbox{$\{\alpha, \beta \} \subset \{ a, b, c, d \}$}, 
we 
need 
to 
know 
how to 
express 
time derivatives of the auxiliary fields. 
Using (\ref{a-system})--(\ref{d-system}), 
we can 
obtain all 
necessary expressions
in 
the 
local forms, 
{\it e.g.},
\begin{align}
& \partial_{t_b}A_{n,m}^{(0)}= 
	-\left( q_{n-1,m+1} r_{n,m} -q_{n-1,m} r_{n,m-1} \right),
\nonumber \\
& \partial_{t_c}A_{n,m}^{(0)}= -\left( C_{n,m+1}^{(0)}-C_{n,m}^{(0)} \right),
\nonumber \\
& \partial_{t_d}A_{n,m}^{(0)}=  q_{n,m+1} D_{n,m+1}^{(0)} r_{n-1,m} 
	- q_{n,m} D_{n,m}^{(0)} r_{n-1,m-1},
\hspace{3mm} \mathrm{etc.}
\nonumber 
\end{align}
Here, we assumed 
that all ``integration constants'' etc.\ 
can be set equal to zero. 
With these local expressions, 
we can check 
the commutativity 
of the four flows 
by direct computations.
Note that 
\mbox{$\lim_{n \to - \infty} \partial_{t_c}A_{n,m}^{(0)}$}, 
\mbox{$\lim_{m \to - \infty} \partial_{t_d}B_{n,m}^{(0)}$}, 
\mbox{$\lim_{n \to - \infty} \partial_{t_a}C_{n,m}^{(0)}$}, and 
\mbox{$\lim_{m \to - \infty} \partial_{t_b}D_{n,m}^{(0)}$} 
do not vanish in general. 
Thus, 
the $t_a$-flow 
can change the boundary value 
of the auxiliary field
in the $t_c$-flow 
and vice versa; the same applies for 
the $t_b$-flow and 
$t_d$-flow. 

\subsection{Bilinearization}

Because the four 
systems 
(\ref{a-system})--(\ref{d-system}) 
are compatible, in the sense that their flows mutually commute, 
we will consider here their common solution 
denoted as 
\mbox{$q_{n,m}(t_a, t_b, t_c, t_d)$}, 
\mbox{$r_{n,m}(t_a, t_b, t_c, t_d)$}, etc.
Here, $t_a$, $t_b$, $t_c$, and $t_d$ are 
independent arguments. 
Thus, 
the solution of 
the 
original system (\ref{DS1}) 
is obtained by setting 
\begin{equation}
\label{t-t}
t_a = at, \;\;\; t_b= bt, \;\;\; t_c= ct, \;\;\; t_d=dt, 
\end{equation}
which indeed 
implies
the relation \mbox{$\partial_t = a\partial_{t_a}+b\partial_{t_b}
+c\partial_{t_c}+d\partial_{t_d}$}. 

We assume 
a solution 
expressible 
in the form, 
\begin{subequations}
\label{bi-trans1}
\begin{align}
& q_{n,m} = \frac{G_{n,m}}{F_{n+1,m}}, \hspace{5mm}
r_{n,m} = \frac{H_{n,m}}{F_{n,m+1}}, \hspace{5mm}
\label{FGH} \\[1mm]
&
A^{(0)}_{n,m}= \partial_{t_a} 
	\log \left( \frac{F_{n,m+1}}{F_{n,m}}\right)
 = \boldsymbol{\Delta}_m^+ \left( 
	\frac{\partial_{t_a} F_{n,m}}{F_{n,m}}\right), 
\label{A0} \\[1mm]
& B^{(0)}_{n,m}=\partial_{t_b} 
	\log \left( \frac{F_{n+1,m}}{F_{n,m}}\right)
 = \boldsymbol{\Delta}_n^+ \left( 
	\frac{\partial_{t_b} F_{n,m}}{F_{n,m}}\right), 
\label{B0} \\[1mm]
& C^{(0)}_{n,m}=\frac{F_{n,m+1}F_{n,m-1}}{\left( F_{n,m}\right)^{2}}, 
\hspace{5mm}
D^{(0)}_{n,m}=\frac{F_{n+1,m}F_{n-1,m}}{\left( F_{n,m}\right)^{2}}, 
\label{C0D0}
\end{align}
\end{subequations}
and bilinearize the four 
systems (\ref{a-system})--(\ref{d-system}) 
in terms of the 
``tau functions'' $F_{n,m}$, $G_{n,m}$, and $H_{n,m}$ 
as follows:
\\
\vspace{-3mm}
\\
$\bullet$ $a$-system
\begin{subequations}
\label{a-system2}
\begin{align}
& F_{n+1,m+1} \partial_{t_a} G_{n,m} -G_{n,m} \partial_{t_a} F_{n+1,m+1} 
 = F_{n+1,m} G_{n,m+1},
\label{F-ta-G}
\\
& F_{n,m}\partial_{t_a} H_{n,m} -H_{n,m}\partial_{t_a} F_{n,m}
  =-F_{n,m+1} H_{n,m-1},
\label{F-ta-H}
\\
& F_{n,m}\partial_{t_a} F_{n+1,m} -F_{n+1,m}\partial_{t_a} F_{n,m}
  =-G_{n,m} H_{n,m-1},
\label{F-ta-F}
\end{align}
\end{subequations}
$\bullet$ $b$-system
\begin{subequations}
\label{b-system2}
\begin{align}
& F_{n,m}\partial_{t_b} G_{n,m} -G_{n,m}\partial_{t_b} F_{n,m}
  =-F_{n+1,m} G_{n-1,m},
\label{F-tb-G}
\\
& F_{n+1,m+1} \partial_{t_b} H_{n,m} -H_{n,m} \partial_{t_b} F_{n+1,m+1} 
 = F_{n,m+1} H_{n+1,m},
\label{F-tb-H}
\\
& F_{n,m}\partial_{t_b} F_{n,m+1} -F_{n,m+1}\partial_{t_b} F_{n,m}
  =-G_{n-1,m} H_{n,m},
\label{F-tb-F}
\end{align}
\end{subequations}
$\bullet$ $c$-system
\begin{subequations}
\label{c-system2}
\begin{align}
& F_{n+1,m}\partial_{t_c} G_{n,m} -G_{n,m}\partial_{t_c} F_{n+1,m}
  =F_{n+1,m+1} G_{n,m-1},
\label{}
\\
& F_{n,m+1} \partial_{t_c} H_{n,m} -H_{n,m} \partial_{t_c} F_{n,m+1} 
 = -F_{n,m} H_{n,m+1},
\label{}
\\
& F_{n+1,m}F_{n,m+1} -F_{n+1,m+1} F_{n,m} = G_{n,m} H_{n,m},
\label{FF-GH1}
\end{align}
\end{subequations}
$\bullet$ $d$-system
\begin{subequations}
\label{d-system2}
\begin{align}
& F_{n+1,m}\partial_{t_d} G_{n,m} -G_{n,m}\partial_{t_d} F_{n+1,m}
  =-F_{n,m} G_{n+1,m},
\label{}
\\
& F_{n,m+1} \partial_{t_d} H_{n,m} -H_{n,m} \partial_{t_d} F_{n,m+1} 
 = F_{n+1,m+1} H_{n-1,m},
\label{}
\\
& F_{n+1,m}F_{n,m+1} -F_{n+1,m+1} F_{n,m} = G_{n,m} H_{n,m}.
\label{FF-GH2}
\end{align}
\end{subequations}
To be 
precise, 
each 
triplet of 
bilinear equations
gives 
a sufficient condition 
for the 
corresponding 
original system. 
Note that for
the $c$-system and 
$d$-system, 
the bilinear forms 
as well as some exact solutions 
were
studied 
in~\cite{Hu06}. 

\subsection{General solution formulas}

Once a 
solution 
of 
the 
bilinear equations 
(\ref{a-system2})--(\ref{d-system2}) is
obtained, 
formula (\ref{bi-trans1}) with (\ref{aux-scale}) and (\ref{t-t}) 
provides the solution of 
the \mbox{$(2+1)$}-dimensional Ablowitz--Ladik lattice (\ref{DS1}). 
We assume that it satisfies the boundary conditions (\ref{ALbc1}).
Thus, by applying 
the
nonlocal
transformation (\ref{trans1})
with (\ref{XY}), 
(\ref{trans2}), (\ref{trans3}), and (\ref{trans4}),
we 
can also obtain 
the solution of the discrete Davey--Stewartson system (\ref{sym-dDS}). 
To evaluate 
the effect of 
this nonlocal transformation, 
we 
use 
(\ref{FGH}) 
and (\ref{FF-GH1}) (or (\ref{FF-GH2})) 
to rewrite 
the 
infinite products as
\begin{align}
&{\displaystyle \prod_{j=-\infty}^{n-1} \sqrt{1- q_{j,m} r_{j,m}}}
={\displaystyle \prod_{j=-\infty}^{n-1}
        \sqrt{\frac{F_{j+1,m+1} F_{j,m}}{F_{j+1,m}F_{j,m+1}}}} 
= {\displaystyle \sqrt{\frac{F_{n,m+1}}{F_{n,m}}
\lim_{j \to -\infty}\frac{F_{j,m}}{F_{j,m+1}}}}
\hspace{1pt},
\nonumber \\[2mm]
& \displaystyle \prod_{k=-\infty}^{m-1} \sqrt{1- q_{n,k} r_{n,k}}
= \displaystyle \prod_{k=-\infty}^{m-1}
        \sqrt{\frac{F_{n+1,k+1} F_{n,k}}{F_{n+1,k}F_{n,k+1}}}
= \displaystyle \sqrt{\frac{F_{n+1,m}}{F_{n,m}}
\lim_{k \to - \infty}\frac{F_{n,k}}{F_{n+1,k}}
}\hspace{1pt}.
\nonumber 
\end{align}
Because we assumed \mbox{$\abs{q_{n,m} r_{n,m}} \ll 1$}, 
the value of 
\mbox{$(F_{n+1,m+1} F_{n,m})/(F_{n+1,m}F_{n,m+1})$} 
is always restricted to 
the neighborhood of 
$1$. 
For simplicity, 
in considering the solution of (\ref{sym-dDS}), 
we also assume that 
$F_{n,m}$ is positive 
(or, at least, \mbox{$\left| \arg F_{n,m}\right|$} is sufficiently small); 
the positivity condition 
\mbox{$F_{n,m}>0$} can fully 
justify the use 
of the formulas for the square root, 
such as 
\mbox{$\sqrt{X^2}=X$} and 
\mbox{$\sqrt{X/Y}=\sqrt{X}/\sqrt{Y}$}. 
%
We set the norming functions 
$h_m$ and $l_n$ 
in (\ref{XY}) as 
\[
h_m = \sqrt{\lim_{j \to -\infty}\frac{F_{j,m}}{F_{j,m+1}}}\hspace{1pt}, 
\hspace{5mm}
l_n= \sqrt{\lim_{k \to - \infty}\frac{F_{n,k}}{F_{n+1,k}}}\hspace{1pt}.
\]
Thus, 
we obtain 
\begin{equation}
\label{XY-local}
X_{n,m}= {\displaystyle \sqrt{\frac{F_{n,m+1}}{F_{n,m}}}}
\hspace{1pt}, \hspace{5mm}
Y_{n,m}={\displaystyle \sqrt{\frac{F_{n+1,m}}{F_{n,m}}}}\hspace{1pt}.
\end{equation}
%
%
After all, 
we can 
express the transformation from 
(\ref{DS1}) to 
(\ref{sym-dDS}) locally 
in terms of the ``tau function'' $F_{n,m}$. 
Combining 
(\ref{trans1}), (\ref{trans2}), (\ref{trans3}), 
(\ref{aux-scale}), 
(\ref{bi-trans1}), and (\ref{XY-local}), 
we arrive at
general solution formulas 
for the discrete Davey--Stewartson system (\ref{sym-dDS}) 
in the form, 
%
\begin{subequations}
\begin{align}
\label{}
& u_{n,m}  = \frac{G_{n,m}}{\sqrt{F_{n+1,m} F_{n,m+1}}}, \hspace{5mm}
v_{n,m} = \frac{H_{n,m}}{\sqrt{F_{n+1,m} F_{n,m+1}}},
\label{FGH2}
\\[2mm] & 
\widetilde{A}_{n,m}=\frac{1}{2}
\left( a \partial_{t_a} -b\partial_{t_b}
-c\partial_{t_c}-d\partial_{t_d} \right)
        \log \left( \frac{F_{n,m+1}}{F_{n,m}}\right), 
\\[2mm] & 
\widetilde{B}_{n,m}=\frac{1}{2}
\left( b\partial_{t_b} -a\partial_{t_a}
-c\partial_{t_c}-d\partial_{t_d} \right)
        \log \left( \frac{F_{n+1,m}}{F_{n,m}}\right), 
\\[2mm] & 
f_{n,m}=  \frac{\sqrt{F_{n,m+2}F_{n,m}}}{F_{n,m+1}}, 
\hspace{5mm}
g_{n,m}= \frac{\sqrt{F_{n+2,m}F_{n,m}}}{F_{n+1,m}}.
\end{align}
\end{subequations}
Here, 
the time variables are set 
as in (\ref{t-t}) 
and 
the auxiliary fields
$\alpha_{n,m}$ and $\beta_{n,m}$ are 
determined from 
$\widetilde{A}_{n,m}$ and $\widetilde{B}_{n,m}$
through (\ref{trans4}). 
Using the bilinear 
equations (\ref{F-ta-F}) and (\ref{F-tb-F}) 
and noting that 
(\ref{symm-5}) and (\ref{symm-6}) are
identities in $a$, $b$, $c$, and $d$, 
we obtain 
compact expressions for $\alpha_{n,m}$ and $\beta_{n,m}$, 
\begin{subequations}
\begin{align}
\label{}
& \alpha_{n,m}= \left( -a \partial_{t_a} +c\partial_{t_c} \right)
        \log \left( \frac{F_{n,m+1}}{F_{n,m}}\right), 
\\[1mm] & 
\beta_{n,m}= \left( -b\partial_{t_b} + d\partial_{t_d} \right)
        \log \left( \frac{F_{n+1,m}}{F_{n,m}}\right), 
\end{align}
\end{subequations}
and 
new bilinear equations,
\begin{align}
& F_{n,m}\partial_{t_c} F_{n+1,m} -F_{n+1,m}\partial_{t_c} F_{n,m}
  =G_{n,m-1} H_{n,m},
\label{new-bi1}
\\
& F_{n,m} \partial_{t_d} F_{n,m+1} -F_{n,m+1} \partial_{t_d} F_{n,m}
 = G_{n,m} H_{n-1,m}.
\label{new-bi2}
\end{align}
Note that (\ref{new-bi1}) and (\ref{new-bi2}) 
fill in the 
piece 
missing in (\ref{a-system2})--(\ref{d-system2}). 
In the next subsection, we 
construct 
common 
solutions to
all these bilinear equations. 

As described below (\ref{sym-dDS}),
when \mbox{$c= -a^\ast$} and \mbox{$d= -b^\ast$}, 
we can impose the 
reduction 
\mbox{$v_{n,m}=\sigma u_{n,m}^{\ast}$} 
with a real constant $\sigma$. 
This reduction can be
realized 
by requiring that 
\mbox{$F_{n,m} >0 $} and \mbox{$H_{n,m} =\sigma G_{n,m}^{\hspace{1pt}\ast}$}.
%

\subsection{Solitons and dromions}

The set of 
bilinear equations 
(\ref{a-system2})--(\ref{d-system2}) 
together with 
(\ref{new-bi1}) and (\ref{new-bi2}) is 
not ideally 
symmetric in its
present form. 
In particular, it is not clear 
why 
reductions such as 
\mbox{$F_{n,m}^{\hspace{1pt}\ast} = F_{n,m}$} 
and \mbox{$H_{n,m} =\sigma G_{n,m}^{\hspace{1pt}\ast}$} 
are allowed. 
To restore the symmetry, we 
need only to 
rewrite 
(\ref{F-ta-G}), (\ref{F-ta-H}), 
(\ref{F-tb-G}), and (\ref{F-tb-H}) 
using (\ref{F-ta-F}), (\ref{F-tb-F}), 
and (\ref{FF-GH1}) 
(or (\ref{FF-GH2})). 
For example, 
using (\ref{F-ta-F}) 
and then (\ref{FF-GH1}), 
(\ref{F-ta-G}) can be rewritten as
\[
 F_{n,m+1} \partial_{t_a} G_{n,m} -G_{n,m} \partial_{t_a} F_{n,m+1}
 = F_{n,m} G_{n,m+1}. 
\]
Thus, the full set of bilinear equations can be 
reformulated in 
the 
symmetric form,
\begin{align}
& F_{n+1,m}F_{n,m+1} -F_{n+1,m+1} F_{n,m} = G_{n,m} H_{n,m},
\hspace{15mm}
\label{bilin1}
\end{align}
\vspace{-8mm}
\begin{subequations}
\label{a-system3}
\begin{align}
& F_{n,m+1} \partial_{t_a} G_{n,m} -G_{n,m} \partial_{t_a} F_{n,m+1}
 = F_{n,m} G_{n,m+1},
\label{bilin2}
\\
& F_{n+1,m}\partial_{t_a} H_{n,m} -H_{n,m}\partial_{t_a} F_{n+1,m}
  =-F_{n+1,m+1} H_{n,m-1},
\label{bilin3}
\\
& F_{n,m}\partial_{t_a} F_{n+1,m} -F_{n+1,m}\partial_{t_a} F_{n,m}
  =-G_{n,m} H_{n,m-1},
\label{bilin4}
\end{align}
\end{subequations}
\vspace{-8mm}
\begin{subequations}
\label{b-system3}
\begin{align}
& F_{n,m+1}\partial_{t_b} G_{n,m} -G_{n,m}\partial_{t_b} F_{n,m+1}
  =-F_{n+1,m+1} G_{n-1,m},
\label{bilin5}
\\
& F_{n+1,m} \partial_{t_b} H_{n,m} -H_{n,m} \partial_{t_b} F_{n+1,m}
 = F_{n,m} H_{n+1,m},
\label{bilin6}
\\
& F_{n,m}\partial_{t_b} F_{n,m+1} -F_{n,m+1}\partial_{t_b} F_{n,m}
  =-G_{n-1,m} H_{n,m},
\label{bilin7}
\end{align}
\end{subequations}
\vspace{-8mm}
\begin{subequations}
\begin{align}
& F_{n+1,m}\partial_{t_c} G_{n,m} -G_{n,m}\partial_{t_c} F_{n+1,m}
  =F_{n+1,m+1} G_{n,m-1},
\label{bilin8}
\\
& F_{n,m+1} \partial_{t_c} H_{n,m} -H_{n,m} \partial_{t_c} F_{n,m+1}
 = -F_{n,m} H_{n,m+1},
\label{bilin9}
\\
& F_{n,m}\partial_{t_c} F_{n+1,m} -F_{n+1,m}\partial_{t_c} F_{n,m}
  =G_{n,m-1} H_{n,m},
\label{bilin10}
\end{align}
\end{subequations}
\vspace{-8mm}
\begin{subequations}
\label{d-system3}
\begin{align}
& F_{n+1,m}\partial_{t_d} G_{n,m} -G_{n,m}\partial_{t_d} F_{n+1,m}
  =-F_{n,m} G_{n+1,m},
\label{bilin11}
\\
& F_{n,m+1} \partial_{t_d} H_{n,m} -H_{n,m} \partial_{t_d} F_{n,m+1}
 = F_{n+1,m+1} H_{n-1,m},
\label{bilin12}
\\
& F_{n,m} \partial_{t_d} F_{n,m+1} -F_{n,m+1} \partial_{t_d} F_{n,m}
 = G_{n,m} H_{n-1,m}.
\label{bilin13}
\end{align}
\end{subequations}
%
It is now clear that the $t_a$-flow and 
$t_b$-flow 
can be 
identified with 
the $t_c$-flow and 
$t_d$-flow, respectively, 
through 
the complex conjugation reduction. 
Moreover, 
(\ref{a-system3}) and (\ref{b-system3}) 
correspond to each other 
by the interchange 
of $n$ and $m$, up to a redefinition of the variables. 
With these symmetries in mind, we can 
considerably 
reduce the task of constructing
explicit solutions 
to 
the above 13
bilinear equations. 

In 
the same way as in 
the continuous case 
(see, {\it e.g.}, 
\cite{Hirota90}), 
we can construct the one-soliton solution and 
a two-soliton solution 
straightforwardly. 
The 
one-soliton solution is given by 
\begin{align}
& F_{n,m} = 1 -\frac{g \widebar{g}}
{(1-p \widebar{p}\hspace{1pt})(1-q \widebar{q}\hspace{1pt})}
(p \widebar{p}\hspace{1pt})^n (q \widebar{q}\hspace{1pt})^m
\mathrm{e}^{\omega +\widebar{\omega}},
\nonumber \\
& G_{n,m}= g \hspace{1pt}p^n q^m \mathrm{e}^{\omega}, \hspace{5mm}
  H_{n,m}= \widebar{g} \hspace{1pt}\widebar{p}\hspace{2pt}^n 
	\widebar{q}\hspace{2pt}^m 
	\mathrm{e}^{\widebar{\omega}},
\nonumber 
\end{align}
where 
\mbox{$\omega:=q \hspace{1pt}t_a -p^{-1}t_b + q^{-1} t_c-p \hspace{1pt}t_d$}, 
\mbox{$\widebar{\omega} := -\widebar{q}\hspace{1pt}^{-1} t_a 
 +\widebar{p}\hspace{2pt}t_b 
 -\widebar{q}\hspace{2pt} t_c+\widebar{p}\hspace{1pt}^{-1}t_d$}, 
and $g$, $p$, $q$, etc.\ are
nonzero 
constants. 
The constant $q$ should not be confused with $q_{n,m}$.
Actually, we can shift 
$t_a$, $t_b$, $t_c$, and $t_d$ 
in (\ref{t-t})
by arbitrary 
constants, 
but this freedom 
can be absorbed by 
rescaling $g$ and $\widebar{g}$. 
When \mbox{$c= -a^\ast$} and \mbox{$d= -b^\ast$} 
(cf.~(\ref{t-t})),
we 
set 
\mbox{$\widebar{p}= p^\ast$}, 
\mbox{$\widebar{q}= q^\ast$}, 
and \mbox{$\widebar{g}= \sigma g^\ast$}
with \mbox{$\sigma 
(1-|p|^2)(1-|q|^2)<0$}. 
Thus, 
\mbox{$F_{n,m} >0 $} and \mbox{$H_{n,m} =\sigma G_{n,m}^{\hspace{1pt}\ast}$}, 
so 
the complex conjugation reduction is realized. 

To save space, 
we omit 
a rather 
lengthy expression for 
a two-soliton solution.
These 
soliton 
solutions 
are direct 
\mbox{$(2+1)$}-dimensional 
analogues 
of 
the soliton solutions 
of the \mbox{$(1+1)$}-dimensional systems 
such as 
(\ref{sd-AL}). 
With an appropriate choice of the parameters, 
they represent
straight 
line solitons 
in the physical variables 
(cf.~(\ref{FGH}) 
or 
(\ref{FGH2})) 
and thus are not 
localized. 

In the following, we 
obtain 
more interesting 
solutions, 
that is, 
dromion solutions;
dromions~\cite{FoSan} 
are 
spatially 
localized 
``solitons''
that 
decay exponentially 
in all 
directions~\cite{Boiti88} 
and 
can 
exhibit 
nontrivial interaction 
properties~\cite{Hirota90,GN91}.
More details as well as 
an extensive list of 
references 
can be found in the review article~\cite{Boiti95}. 
In analogy with the continuous case~\cite{Hirota90}, 
the 
one-dromion solution is
obtained as 
\begin{align}
 F_{n,m} & = 1 +\frac{\alpha_{11}}{1-p \widebar{p}\hspace{1pt}}
 (p \widebar{p}\hspace{1pt})^n\mathrm{e}^{\omega_1 +\widebar{\omega}_1}
+\frac{\alpha_{22}}{1-q \widebar{q}\hspace{1pt}}
 (q \widebar{q}\hspace{1pt})^m \mathrm{e}^{\omega_2 +\widebar{\omega}_2}
\nonumber \\
& \hphantom{=}\; + \frac{\alpha_{11}\alpha_{22}-\alpha_{12}\alpha_{21}}
{(1-p \widebar{p}\hspace{1pt})(1-q \widebar{q}\hspace{1pt})}
(p \widebar{p}\hspace{1pt})^n (q \widebar{q}\hspace{1pt})^m 
 \mathrm{e}^{\omega_1 +\omega_2 +\widebar{\omega}_1+\widebar{\omega}_2},
\nonumber \\[2mm]
 G_{n,m} &= \alpha_{12} 
p^n q^m \mathrm{e}^{\omega_1+\omega_2}, \hspace{5mm}
  H_{n,m}= \alpha_{21}\widebar{p}\hspace{2pt}^n 
	\widebar{q}\hspace{2pt}^m 
	\mathrm{e}^{\widebar{\omega}_1+\widebar{\omega}_2},
\nonumber 
\end{align}
where 
\[
\omega_1:= -p^{-1}t_b -p \hspace{1pt}t_d, \hspace{5mm}
\widebar{\omega}_1 := \widebar{p}\hspace{2pt}t_b 
	+\widebar{p}\hspace{2pt}^{-1}t_d, \hspace{5mm}
\omega_2 :=q \hspace{1pt}t_a + q^{-1} t_c, \hspace{5mm}
\widebar{\omega}_2 := 
 -\widebar{q}\hspace{2pt}^{-1} t_a -\widebar{q}\hspace{2pt} t_c.
\]
When \mbox{$c= -a^\ast$} and \mbox{$d= -b^\ast$} 
(cf.~(\ref{t-t})),
we
set 
\mbox{$\widebar{p}= p^\ast$}, 
\mbox{$\widebar{q}= q^\ast$}, 
and \mbox{$\alpha_{21}= \sigma \alpha_{12}^\ast$} 
so that 
\mbox{$H_{n,m} =\sigma G_{n,m}^{\hspace{1pt}\ast}$}.
Moreover, 
if 
the 
constant 
coefficients 
of the three 
terms 
in $F_{n,m}$ 
are 
positive, then 
\mbox{$F_{n,m} >0 $}, 
so 
the complex conjugation reduction is realized. 
Note that 
the above 
$F_{n,m}$ 
can be written 
in a \mbox{$2 \times 2$} determinant form, 
\begin{align}
 F_{n,m} & = 
\det \left\{
I + 
\left(
\begin{array}{cc}
\alpha_{11} & \alpha_{21} \\
\alpha_{12} & \alpha_{22} \\
\end{array}
\right)
\left(
\begin{array}{cc}
p^n\mathrm{e}^{\omega_1} & \\
 &  \widebar{q}\hspace{2pt}^m \mathrm{e}^{\widebar{\omega}_2}\\
\end{array}
\right)
\left(
\begin{array}{cc}
\frac{1}{1-p \widebar{p}\hspace{1pt}} &  \\
 & \frac{1}{1-q \widebar{q}\hspace{1pt}} \\
\end{array}
\right)
\left(
\begin{array}{cc}
\widebar{p}\hspace{2pt}^n \mathrm{e}^{\widebar{\omega}_1} & \\
 & q^m \mathrm{e}^{\omega_2} \\
\end{array}
\right)
\right\}.
\nonumber 
\end{align}

Following the Gilson--Nimmo approach~\cite{GN91} 
in the continuous case, 
we construct the multidromion solution called 
the 
\mbox{$(M,N)$}-dromion solution. 
The one-dromion solution corresponds 
to the simplest case of \mbox{$M=N=1$}. 
%
%
Hereinafter, 
we 
suppress 
the subscripts 
of the 
functions 
representing 
their 
dependence on
the spatial variables $n$ and $m$. 
When 
they 
are shifted, we express it using 
the shift operators 
\[
(\vt{S}_n Z)_{n,m} := Z_{n+1,m}, \hspace{5mm}
(\vt{S}_m Z)_{n,m} := Z_{n,m+1}.
\]
We will consider 
various \mbox{$(M+N) \times (M+N)$} 
matrices; they 
all have the same shape as 
\mbox{$2 \times 2$} block matrices, 
so 
operations can be performed blockwise. 
%
For simplicity,
off-diagonal zeros in 
the block diagonal matrices 
are omitted. 
We 
introduce two
diagonal matrices 
as
\begin{align}
\nonumber 
& \Xi := 
\left(
\begin{array}{cc}
\Xi_1 & \\
 & \Xi_2 \\
\end{array}
\right), 
\hspace{5mm}
\Xi_1 := \mathrm{diag}(\varphi_1, \ldots, \varphi_M), \hspace{5mm}
\Xi_2 := \mathrm{diag}(\chi_1, \ldots, \chi_N), 
\\[2mm]
& \nonumber 
\widebar{\Xi} := 
\left(
\begin{array}{cc}
\widebar{\Xi}_1 & \\
 & \widebar{\Xi}_2 \\
\end{array}
\right), 
\hspace{5mm}
\widebar{\Xi}_1 := \mathrm{diag}(\widebar{\varphi}_1, \ldots, 
	\widebar{\varphi}_M), \hspace{5mm}
\widebar{\Xi}_2 := \mathrm{diag}(\widebar{\chi}_1, 
	\ldots, \widebar{\chi}_N), 
\end{align}
where
\begin{align}
& \varphi_i(n) := p_i^n \mathrm{e}^{-p_i^{-1} t_b -p_i t_d}, \hspace{5mm}
\chi_i(m) := q_i^m \mathrm{e}^{-q_i^{-1} t_a -q_i t_c}, 
\nonumber \\[1mm]
& \widebar{\varphi}_i(n) := \widebar{p}_i^{\hspace{2pt}n} 
 \mathrm{e}^{\widebar{p}_i t_b 
	+\widebar{p}_i^{\hspace{1pt}-1} t_d}, \hspace{5mm}
\widebar{\chi}_i(m) := \widebar{q}_i^{\hspace{2pt}m}
\mathrm{e}^{\widebar{q}_i t_a + \widebar{q}_i^{\hspace{1pt}-1} t_c}. 
\nonumber 
\end{align}
We 
also 
introduce 
\begin{align}
\nonumber 
& R := 
\left(
\begin{array}{cc}
P & \\
 & Q \\
\end{array}
\right), 
\hspace{5mm}
P := \mathrm{diag}(p_1, \ldots, p_M), \hspace{5mm}
Q := \mathrm{diag}(q_1, \ldots, q_N), 
\\[2mm]
& \nonumber 
\widebar{R} := 
\left(
\begin{array}{cc}
\widebar{P} & \\
 & \widebar{Q} \\
\end{array}
\right), 
\hspace{5mm}
\widebar{P} := \mathrm{diag}(\widebar{p}_1, \ldots, 
 \widebar{p}_M), \hspace{5mm}
\widebar{Q} := \mathrm{diag}(\widebar{q}_1, \ldots, \widebar{q}_N).
\end{align}
%
Then, the following relations hold:
%
\begin{align}
& \vt{S}_n \Xi = 
\Xi \left(
\begin{array}{cc}
P & \\
 & I\\
\end{array}
\right),
\hspace{5mm}
\vt{S}_m \Xi = \Xi \left(
\begin{array}{cc}
I & \\
 & Q\\
\end{array}
\right),
\label{} 
\nonumber\\[2mm]
& \vt{S}_n \widebar{\Xi} = 
\left(
\begin{array}{cc}
\widebar{P} & \\
 & I\\
\end{array}
\right) \widebar{\Xi},
\hspace{5mm}
\vt{S}_m \widebar{\Xi} = \left(
\begin{array}{cc}
I & \\
 & \widebar{Q} \\
\end{array}
\right) \widebar{\Xi},
\nonumber\\[2mm]
& 
\partial_{t_a} 
\Xi = -\Xi \left(
\begin{array}{cc}
O & \\
 & Q^{-1} \\
\end{array}
\right), \hspace{5mm}
\partial_{t_b} 
\Xi = - \Xi \left(
\begin{array}{cc}
P^{-1} & \\
 & O \\
\end{array}
\right),\hspace{5mm}
\nonumber\\[2mm]
& \partial_{t_c} \Xi 
= -\Xi\left(
\begin{array}{cc}
O & \\
 & Q \\
\end{array}\right),\hspace{5mm}
\partial_{t_d} 
\Xi = -\Xi\left(
\begin{array}{cc}
P & \\
 & O \\
\end{array}
\right),
\nonumber\\[2mm]
& 
\partial_{t_a} 
\widebar{\Xi} = \left(
\begin{array}{cc}
O & \\
 & \widebar{Q}\\
\end{array}
\right) \widebar{\Xi}, \hspace{5mm}
\partial_{t_b} 
\widebar{\Xi} = \left(
\begin{array}{cc}
\widebar{P} & \\
 & O \\
\end{array}
\right) \widebar{\Xi},\hspace{5mm}
\nonumber\\[2mm]
& \partial_{t_c} 
\widebar{\Xi} = \left(
\begin{array}{cc}
O & \\
 & \widebar{Q}^{\hspace{1pt}-1} \\
\end{array}
\right) \widebar{\Xi},\hspace{5mm}
\partial_{t_d} 
\widebar{\Xi} = \left(
\begin{array}{cc}
\widebar{P}^{\hspace{1pt}-1} & \\
 & O \\
\end{array}
\right) \widebar{\Xi}.
\nonumber
\end{align}
Note that the order 
of 
two diagonal matrices 
on the right-hand side 
can be changed 
because they commute.
Moreover, we
introduce 
\mbox{$M \times M $} matrices $K$ and $E_M$ 
as
%
\begin{align}
& (K)_{ij} := \frac{1}{1-p_i \widebar{p}_j}, 
\hspace{5mm}
 (E_M)_{ij} := 1, \hspace{5mm} 1 \le i,j \le M, 
\nonumber 
\end{align}
and \mbox{$N \times N $} matrices $L$ and $E_N$ as 
\begin{align}
& (L)_{kl} := \frac{1}{1-q_k \widebar{q}_{\hspace{1pt}l}}, 
\hspace{5mm}
 (E_N)_{kl} := 1, \hspace{5mm} 1 \le k,l \le N. 
\nonumber 
\end{align}
They satisfy the relations,
\[
 K - P K \widebar{P} =E_M, \hspace{5mm}
 L - Q L \widebar{Q} =E_N.
\]
We 
define 
\mbox{$(M+N)$}-component 
column 
vectors 
as 
\begin{align}
\nonumber 
\vt{e}_M := ( 1, \ldots, 1, 0, \ldots, 0 )^T, 
\hspace{5mm}
\vt{e}_N := (0, \ldots, 0, 1, \ldots, 1)^T,
\end{align}
and 
\begin{align}
\nonumber 
& \vt{l} := \Xi \hspace{1pt}\vt{e}_M
=( \varphi_1, \ldots, \varphi_M, 0, \ldots, 0 )^T, 
\hspace{5mm}
\vt{m} := \Xi \hspace{1pt}\vt{e}_N
 = ( 0, \ldots, 0, \chi_1, \ldots, \chi_N )^T, 
\\[2mm]
& 
\widebar{\vt{l}} := \widebar{\Xi}\hspace{1pt} \vt{e}_M
 =( \widebar{\varphi}_1, \ldots, \widebar{\varphi}_M, 0, \ldots, 0 )^T, 
\hspace{5mm}
\widebar{\vt{m}} := \widebar{\Xi} \hspace{1pt}\vt{e}_N
 = ( 0, \ldots, 0, \widebar{\chi}_1, \ldots, \widebar{\chi}_N )^T.
\nonumber 
\end{align}
Then, we can easily show the following relations: 
\begin{align}
\nonumber 
& \vt{S}_n \hspace{1pt}\vt{l} = R\hspace{2pt}\vt{l} , \hspace{5mm}
\vt{S}_m \hspace{1pt}\vt{l} = \vt{l} , \hspace{5mm}
\vt{S}_n \vt{m} = \vt{m}, \hspace{5mm}
\vt{S}_m \vt{m} = R \hspace{1pt}\vt{m}, 
\\[2mm]
& \vt{S}_n \hspace{1pt}\widebar{\vt{l}} 
 = \widebar{R}\hspace{2pt}\widebar{\vt{l}} , \hspace{5mm}
\vt{S}_m \hspace{1pt}\widebar{\vt{l}} = \widebar{\vt{l}}, \hspace{5mm}
\vt{S}_n \widebar{\vt{m}} = \widebar{\vt{m}}, \hspace{5mm}
\vt{S}_m \widebar{\vt{m}} = \widebar{R} \hspace{1pt}\widebar{\vt{m}}. 
\nonumber 
\end{align}

We set
the ``tau function'' $F_{n,m}$ as
\begin{align}
F 
= \det {\cal F}.
\nonumber 
\end{align}
Here, the \mbox{$(M+N) \times (M+N)$} matrix 
${\cal F}$ is defined as
\begin{align}
{\cal F} := I + {\cal A}\hspace{2pt} \Xi 
\left(
\begin{array}{cc}
 K & \\
 & L \\
\end{array}
\right) 
\widebar{\Xi},
\nonumber 
\end{align}
where ${\cal A}$ is a constant \mbox{$(M+N) \times (M+N)$} matrix. 
We also set
the other 
``tau functions'' $G_{n,m}$ 
and $H_{n,m}$ as
\begin{align}
G 
= \left( \widebar{\vt{m}}^{T} {\cal F}^{-1} {\cal A} 
	\hspace{2pt}\vt{l} \right) F, \hspace{5mm}
H 
= \left(\hspace{1pt} \widebar{\vt{l}}^{\hspace{2pt}T} 
	{\cal F}^{-1} {\cal A} \hspace{1pt}\vt{m} \right) F. 
\nonumber 
\end{align}
Note that 
the 
one-dromion solution
is 
reproduced 
by setting \mbox{$M=N=1$}, up to 
a minor redefinition of the parameters.
Let us check 
that these ``tau functions'' indeed 
satisfy
the only bilinear equation without time 
derivatives, 
(\ref{bilin1}). 
With the aid of the previous relations, we have
\begin{align}
\vt{S}_n {\cal F} -{\cal F} & = {\cal A} \hspace{2pt}\Xi 
\left(
\begin{array}{cc}
P & \\
 & I \\
\end{array}
\right)
\left(
\begin{array}{cc}
K & \\
 & L \\
\end{array}
\right)
\left(
\begin{array}{cc}
\widebar{P} & \\
 & I \\
\end{array}
\right) 
\widebar{\Xi}
-{\cal A} \hspace{2pt}\Xi 
\left(
\begin{array}{cc}
K & \\
 & L \\
\end{array}
\right) \widebar{\Xi}
\nonumber \\
&= -{\cal A} \hspace{2pt}\Xi 
\left(
\begin{array}{cc}
K-P K \widebar{P} & \\
 & O \\
\end{array}
\right) \widebar{\Xi}
\nonumber \\
&= -{\cal A} \hspace{2pt}\Xi \hspace{1pt}\vt{e}_M \hspace{1pt}
 \vt{e}_M^{\hspace{1pt}T} \hspace{1pt}\widebar{\Xi}
\nonumber \\
&=-{\cal A} \hspace{2pt}  \vt{l} \hspace{2pt}\widebar{\vt{l}}^{\hspace{2pt}T}.
\nonumber 
\end{align}
Thus, we obtain 
\begin{align}
\vt{S}_n  F &= \det \left( {\cal F} -{\cal A} \hspace{2pt}  
	\vt{l} \hspace{2pt}\widebar{\vt{l}}^{\hspace{2pt}T}\right)
\nonumber \\
&= \det \left( I - {\cal F}^{-1} {\cal A} \hspace{2pt}  
	\vt{l} \hspace{2pt}\widebar{\vt{l}}^{\hspace{2pt}T}\right) F
\nonumber \\
&= \left( 1 - \widebar{\vt{l}}^{\hspace{2pt}T}
 {\cal F}^{-1} {\cal A} \hspace{2pt} \vt{l} \right) F.
\nonumber 
\end{align}
Similarly, we 
obtain 
\begin{align}
\vt{S}_m  F &= \left( 1 - \widebar{\vt{m}}^{T}
 {\cal F}^{-1} {\cal A} \hspace{1pt} \vt{m} \right) F.
\nonumber 
\end{align}
To 
compute 
\mbox{$\vt{S}_n \vt{S}_m  F$}, we still need to know
\mbox{$\vt{S}_n {\cal F}^{-1}$}. 
This can be expressed 
as 
\begin{align}
\vt{S}_n {\cal F}^{-1} &= \left( \vt{S}_n {\cal F} \right)^{-1} 
\nonumber \\
&= \left( {\cal F} -{\cal A} \hspace{2pt} 
	\vt{l} \hspace{2pt}\widebar{\vt{l}}^{\hspace{2pt}T}\right)^{-1}
\nonumber \\
&= {\cal F}^{-1} + \frac{{\cal F}^{-1}{\cal A} \hspace{2pt} 
	\vt{l} \hspace{2pt}\widebar{\vt{l}}^{\hspace{2pt}T}{\cal F}^{-1}}
{1- \widebar{\vt{l}}^{\hspace{2pt}T} {\cal F}^{-1}{\cal A} \hspace{2pt}\vt{l}}.
\nonumber 
\end{align}
Here, 
we 
used the so-called Sherman--Morrison formula; 
recall 
that 
\mbox{${\cal A} \hspace{2pt} \vt{l}$} is a column vector and 
\mbox{$\widebar{\vt{l}}^{\hspace{2pt}T}$} is a row vector. 
Combining the above results, we obtain 
\begin{align}
\left( \vt{S}_n \vt{S}_m  F \right) F
&=
\left[ 1 - \widebar{\vt{m}}^{T}
 \left( \vt{S}_n {\cal F}^{-1}\right) 
 {\cal A} \hspace{1pt} \vt{m} \right] 
\left(\vt{S}_n F \right) F 
\nonumber \\[1mm]
&= \left(\vt{S}_n F \right) 
\left( 1 - \widebar{\vt{m}}^{T}
 {\cal F}^{-1} {\cal A} \hspace{1pt} \vt{m} \right) F 
 -\left( \widebar{\vt{m}}^{T} {\cal F}^{-1} {\cal A}
        \hspace{2pt}\vt{l} \right) F 
 \left(\hspace{1pt} \widebar{\vt{l}}^{\hspace{2pt}T}
        {\cal F}^{-1} {\cal A} \hspace{1pt}\vt{m} \right) F 
\nonumber 
\nonumber \\[1mm]
&= \left(\vt{S}_n F \right) \left(\vt{S}_m F \right) -GH. 
\nonumber 
\end{align}
This completes the proof of (\ref{bilin1}). 

It is a direct but lengthy calculation to check the 
remaining 
12 equations 
(\ref{bilin2})--(\ref{bilin13}) 
involving time derivatives. 
For example, 
in order 
to check (\ref{bilin3}), 
we need the following intermediate formulas: 
\begin{align}
\partial_{t_a} F &= 
 -\left( \widebar{\vt{m}}^{T}
 {\cal F}^{-1} {\cal A}\hspace{1pt} R^{-1}\vt{m} \right) F, 
\nonumber \\[1mm]
\partial_{t_a} H &= 
\left( \widebar{\vt{l}}^{\hspace{2pt}T}
 {\cal F}^{-1} {\cal A}\hspace{1pt} R^{-1}\vt{m} \right) 
\left( \widebar{\vt{m}}^{T}
 {\cal F}^{-1} {\cal A} \hspace{1pt} \vt{m} -1 \right) F
 - \left(\hspace{1pt} \widebar{\vt{l}}^{\hspace{2pt}T}
        {\cal F}^{-1} {\cal A} \hspace{1pt}\vt{m} \right) 
 \left( \widebar{\vt{m}}^{T}
 {\cal F}^{-1} {\cal A}\hspace{1pt} R^{-1}\vt{m} \right) F, 
\nonumber
\nonumber \\[1mm]
\vt{S}_m^{-1} H 
&=  \left( \widebar{\vt{l}}^{\hspace{2pt}T}
 {\cal F}^{-1} {\cal A}\hspace{1pt} R^{-1}\vt{m} \right) F.
\nonumber
\end{align}
To 
obtain
the first 
formula, we first 
compute \mbox{$\partial_{t_a} {\cal F}$} as 
\begin{align}
\partial_{t_a} {\cal F} & 
= - {\cal A} \hspace{2pt}\Xi
\left(
\begin{array}{cc}
O & \\
 & Q^{-1} \\
\end{array}
\right)
\left(
\begin{array}{cc}
K & \\
 & L \\
\end{array}
\right)
\widebar{\Xi}
+ {\cal A} \hspace{2pt}\Xi
\left(
\begin{array}{cc}
K & \\
 & L \\
\end{array}
\right) 
\left(
\begin{array}{cc}
O & \\
 & \widebar{Q}\\
\end{array}
\right)
\widebar{\Xi}
\nonumber \\
&= -{\cal A} \hspace{2pt}\Xi
\left(
\begin{array}{cc}
O & \\
 & Q^{-1} E_N \\
\end{array}
\right) \widebar{\Xi}
\nonumber \\
&= -{\cal A} \hspace{1pt}R^{-1} \Xi \hspace{1pt}
 \vt{e}_N \hspace{1pt}
 \vt{e}_N^{\hspace{1pt}T} \hspace{1pt}\widebar{\Xi}
\nonumber \\
&=-{\cal A} \hspace{1pt}R^{-1}
	\vt{m} \hspace{1pt}\widebar{\vt{m}}^{T}. 
\nonumber
\end{align}
Then, we multiply it
by 
${\cal F}^{-1}$ 
and take the trace. 
%
%
%

\section{Concluding remarks}

In this paper, we 
have studied 
a suitable space discretization 
of the Davey--Stewartson system. 
The Davey--Stewartson system is 
an 
integrable 
NLS system in \mbox{$2+1$} dimensions, 
which involves 
two spatial variables 
on an equal footing
and allows the complex conjugation reduction 
between the 
dependent 
variables. 
We started with 
a natural 
\mbox{$(2+1)$}-dimensional generalization 
of the Ablowitz--Ladik lattice
and then 
considered 
a nonlocal 
change 
of dependent variables 
to symmetrize the equations of motion. 
Consequently, 
we 
obtained
the space-discrete Davey--Stewartson system 
inheriting
most of the 
important properties of the continuous system; 
in particular, 
it is integrable and 
allows the complex conjugation reduction. 
The price 
to pay 
is 
the irrationality 
of 
the equations of motion 
and their high degree of nonlocality, 
which 
are not seen
in the continuous 
case. 
Through 
a simple reduction, 
we 
reduced the degree of nonlocality 
and obtained a discrete 
modified KdV-type system 
in 
two spatial
dimensions, 
namely, 
the \mbox{$(2+1)$}-dimensional 
modified Volterra lattice (\ref{2DmLV}). 

The \mbox{$(2+1)$}-dimensional Ablowitz--Ladik lattice, 
as well as  the space-discrete Davey--Stewartson system, 
is a superposition of four elementary 
flows that are mutually commutative. 
Naturally, 
the number of 
elementary flows 
is 
equal to the 
number of
directions 
on the square lattice.
Note also 
that 
both 
the \mbox{$(1+1)$}-dimensional Ablowitz--Ladik lattice 
and the continuous Davey--Stewartson system 
can be written as a sum of 
two 
commuting flows. 
We conjecture that 
the \mbox{$(2+1)$}-dimensional Ablowitz--Ladik lattice 
and the space-discrete Davey--Stewartson system 
possess four infinite 
sets 
of higher 
symmetries. 
As in the original 
Ablowitz--Ladik lattice~\cite{Vek02,Sada},
each 
set of 
symmetries
could be 
generated from a single 
discrete-time system 
using the Maclaurin expansion in 
the step-size
parameter. 
It would be 
interesting 
to 
provide
a more precise description 
within 
the framework of 
the 
Sato theory, {\it e.g.}, 
the discrete two-component KP hierarchy (cf.~\cite{Hu07}).

Using the Hirota bilinear method, 
we 
have constructed 
exact solutions such as 
the multidromion solutions 
of the \mbox{$(2+1)$}-dimensional Ablowitz--Ladik lattice 
and the space-discrete Davey--Stewartson system
concurrently. 
Their solutions can be 
obtained 
from 
the same set of bilinear equations, 
although 
their bilinearizing transformations 
are 
rather different 
(cf.~(\ref{FGH}) and (\ref{FGH2})). 
Note that (\ref{FGH2}) 
reflects the 
irrationality
of 
the space-discrete Davey--Stewartson system 
that can, however, allow the complex conjugation reduction. 
%
The solutions 
were derived as the common solutions 
of the four 
elementary 
flows. 
On the level of the bilinear equations, 
the 
four flows look fully 
symmetric 
and stand on an equal footing. 
Thus, it is relatively easy to 
construct their common solutions 
despite the high number of 
bilinear 
equations. 


\section*{Acknowledgments}
The authors thank
Professor Folkert M\"uller-Hoissen
and
Dr.\ Ken-ichi
Maruno
for their useful comments. 


\addcontentsline{toc}{section}{References}
\begin{thebibliography}{99}

\bibitem{GGKM}
C.\ S.\ Gardner, J.\ M.\ Greene, M.\ D.\ Kruskal and R.\ M.\ Miura:
{\em Method for solving the Korteweg--de Vries equation},
Phys.\ Rev.\ Lett.\ {\bf 19} (1967) 1095--1097.

\bibitem{Lax}
P.\ D.\ Lax:
{\em Integrals of nonlinear equations of evolution and solitary waves}, 
Commun.\ Pure Appl.\ Math.\ {\bf 21} (1968) 467--490.

\bibitem{Suris03}
Y.\ B.\ Suris:
{\it The Problem of Integrable Discretization:\ Hamiltonian Approach}
(Birkh\"auser, Basel, 2003).

\bibitem{AKNS}
M.\ J.\ Ablowitz, D.\ J.\ Kaup, A.\ C.\ Newell and H.\ Segur:
{\em Nonlinear-evolution equations of physical significance},
Phys.\ Rev.\ Lett.\ {\bf 31} (1973) 125--127.

\bibitem{ZS1}
V.\ E.\ Zakharov and A.\ B.\ Shabat: 
{\em Exact theory of two-dimensional self-focusing and 
one-dimensional self-modulation of waves in nonlinear media}, 
Sov.\ Phys.--JETP {\bf 34} (1972) 62--69.

\bibitem{ZS2}
V.\ E.\ Zakharov and A.\ B.\ Shabat: 
{\em Interaction between solitons in a stable medium},
Sov.\ Phys.--JETP {\bf 37} (1973) 823--828.

\bibitem{AL1}
M.\ J.\ Ablowitz and J.\ F.\ Ladik:
{\em Nonlinear differential--difference equations and Fourier analysis},
J.\ Math.\ Phys.\ {\bf 17} (1976) 1011--1018.

\bibitem{Tsuchi02}
T.\ Tsuchida: 
{\em Integrable discretizations of 
derivative nonlinear Schr\"odinger equations},
J.\ Phys.\ A:\ Math.\ Gen.\ {\bf 35} (2002) 7827--7847.

\bibitem{2010JPA}
T.\ Tsuchida: 
{\em A systematic method for constructing time
discretizations of integrable lattice systems:\ local
equations of motion}, 
J.\ Phys.\ A:\ Math.\ Theor.\ {\bf 43} (2010) 415202 (22pp).

\bibitem{DS74}
A.\ Davey and K.\ Stewartson: 
{\em On three-dimensional packets of surface waves}, 
Proc.\ R.\ Soc.\ Lond.\ A {\bf 338} (1974) 101--110. 

\bibitem{Benney69} 
D.\ J.\ Benney and G.\ J.\ Roskes:
{\em Wave instabilities}, 
Stud.\ Appl.\ Math.\ {\bf 48} (1969) 377--385.

\bibitem{Hab75} 
M.\ J.\ Ablowitz and R.\ Haberman: 
{\em Nonlinear evolution equations---two and three dimensions}, 
Phys.\ Rev.\ Lett.\ {\bf 35} (1975) 1185--1188. 

\bibitem{Morris77}
H.\ C.\ Morris: 
{\em Prolongation structures and nonlinear evolution equations 
in two spatial dimensions.\ II.\ 
A generalized nonlinear Schr\"odinger equation}, 
J.\ Math.\ Phys.\ {\bf 18} (1977) 285--288.

\bibitem{Ab78}
M.\ J.\ Ablowitz: 
{\em Lectures on the inverse scattering transform},
Stud.\ Appl.\ Math.\ {\bf 58} (1978) 17--94.

\bibitem{Anker} 
D.\ Anker and N.\ C.\ Freeman: 
{\em On the soliton solutions of the Davey--Stewartson equation for 
long waves}, 
Proc.\ R.\ Soc.\ Lond.\ A {\bf 360} (1978) 529--540.

\bibitem{Cornille}
H. Cornille: 
{\em Solutions of the generalized nonlinear Schr\"odinger equation 
in two spatial dimensions}, J.\ Math.\ Phys.\ {\bf 20} (1979) 199--209.


\bibitem{Calo76}
F.\ Calogero and A.\ Degasperis:
{\em Nonlinear evolution equations solvable by the inverse
spectral transform.\ I}, 
Nuovo Cimento B {\bf 32} (1976) 201--242.

\bibitem{Hu06} 
Gegenhasi, 
X.-B.\
Hu and D.\ Levi: 
{\em On a discrete Davey--Stewartson system}, 
Inverse Probl.\ {\bf 22} (2006) 1677--1688.

\bibitem{Hu07} 
Gegenhasi, 
X.-B.\ 
Hu, D.\ Levi and S.\ Tsujimoto:
{\em A difference analogue of the Davey--Stewartson
system:\ discrete Gram-type determinant solution
and Lax pair},
J.\ Phys.\ A:\ Math.\ Theor.\ {\bf 40} (2007) 12741--12751.

\bibitem{Hirota04}
R.\ Hirota:
{\it The Direct Method in Soliton Theory}
(Cambridge Univ.\ Press, Cambridge, 2004)
edited and translated by A.\ Nagai, J.\ Nimmo and C.\ Gilson.

\bibitem{AL77}
M.\ J.\ Ablowitz and J.\ F.\ Ladik:
{\em On the solution of a class of nonlinear partial difference equations},
Stud.\ Appl.\ Math.\ {\bf 57} (1977) 1--12.

\bibitem{Chiu77}
S.-C.\ Chiu and J.\ F.\ Ladik: 
{\em Generating exactly soluble nonlinear discrete evolution
 equations by a generalized Wronskian technique}, 
J.\ Math.\ Phys.\ {\bf 18} (1977) 690--700.

\bibitem{Nizh80}
L.\ P.\ Nizhnik: 
{\em Integration of multidimensional nonlinear equations 
by the method of the inverse problem},
Sov.\ Phys.\ Dokl.\ {\bf 25} (1980) 706--708. 

\bibitem{Nizh82}
L.\ P.\ Nizhnik and M.\ D.\ Pochinaiko:
{\em Integration of the nonlinear two-dimensional spatial 
Schr\"odinger equation by the inverse-problem method},
Funct.\ Anal.\ Appl.\ {\bf 16} (1982) 66--69.

\bibitem{Kaji90}
K.\ Kajiwara, J.\ Matsukidaira and J.\ Satsuma: 
{\em Conserved quantities of two-component KP hierarchy}, 
Phys.\ Lett.\ A {\bf 146} (1990) 115--118.

\bibitem{MY97}
A.\ V.\ Mikhailov and R.\ I.\ Yamilov:
{\em On integrable two-dimensional generalizations of 
nonlinear Schr\"odinger type equations}, 
Phys.\ Lett.\ A {\bf 230} (1997) 295--300.

\bibitem{Kako}
F.\ Kako and N.\ Mugibayashi: 
{\em Complete integrability of 
general nonlinear differential-difference equations 
solvable by the inverse method.\ II}, 
Prog.\ Theor.\ Phys.\ {\bf 61} (1979) 776--790.

\bibitem{Kulish}
P.\ P.\ Kulish: 
{\em Quantum difference nonlinear Schr\"odinger equation}, 
Lett.\ Math.\ Phys.\ 
{\bf 5} (1981) 191--197.

\bibitem{GIK84}
V.\ S.\ Gerdjikov, M.\ I.\ Ivanov and P.\ P.\ Kulish:
{\em Expansions over the ``squared'' solutions and difference
evolution equations}, J.\ Math.\ Phys.\ {\bf 25} (1984) 25--34.

\bibitem{GI2}
V.\ S.\ Gerdzhikov and  M.\ I.\ Ivanov:
{\em Hamiltonian structure of multicomponent nonlinear
Schr\"{o}dinger equations in difference form},
Theor.\ Math.\ Phys.\ {\bf 52} (1982) 676--685.

\bibitem{Dmitry09-1} 
D.\ 
Zakharov: 
{\em A discrete analogue of the Dirac operator and the discrete
modified Novikov--Veselov hierarchy},
Int.\ Math.\ Res.\ Not.\ 
{\bf 2010} 
No.~18 
3463--3488. 

\bibitem{Dmitry09-2} 
D.\ V.\ Zakharov: 
{\em  
Weierstrass representation 
for discrete isotropic surfaces 
in $\mathbb{R}^{2,1}$, $\mathbb{R}^{3,1}$, and $\mathbb{R}^{2,2}$}, 
Funct.\ Anal.\ Appl.\ {\bf 45} (2011) 25--32.

\bibitem{Bogdanov1}
L.\ V.\ Bogdanov: 
{\em Veselov--Novikov equation as a natural two-dimensional 
generalization of the Korteweg--de Vries equation},
Theor.\ Math.\ Phys.\ {\bf 70} (1987) 219--223. 

\bibitem{Bogdanov2}
L.\ V.\ Bogdanov: 
{\em On the two-dimensional Zakharov--Shabat problem},
Theor.\ Math.\ Phys.\ {\bf 72} (1987) 790--793. 

\bibitem{Boiti88}
M.\ Boiti, J.\ J.\ -P.\ Leon, L.\ Martina and 
F.\ Pempinelli: 
{\em Scattering of localized solitons in the plane},
Phys.\ Lett.\ A {\bf 132} (1988) 432--439.

\bibitem{Hiro73}
R.\ Hirota: 
{\em Exact $N$-soliton solution of 
nonlinear lumped self-dual network equations}, 
J.\ Phys.\ Soc.\ Jpn.\ {\bf 35} (1973) 289--294.

\bibitem{HuJPA05}
X.-B.\ Hu, C.-X.\ Li, J.\ J.\ C.\ Nimmo and G.-F.\ Yu: 
{\em An integrable symmetric \mbox{$(2+1)$}-dimensional Lotka--Volterra 
equation and a family of its solutions}, 
J.\ Phys.\ A:\ Math.\ Gen.\ {\bf 38} (2005) 195--204.

\bibitem{HuJMAA05}
C.-X.\ Li,  J.\ J.\ C.\ Nimmo, X.-B.\ Hu and Gegenhasi: 
{\em On an integrable modified \mbox{$(2+1)$}-dimensional Lotka--Volterra equation}, 
J.\ Math.\ Anal.\ Appl.\ {\bf 309} (2005) 686--700.

\bibitem{Kri10}
S.\ Grushevsky and I.\ Krichever: 
{\em Integrable discrete Schr\"odinger
equations and a characterization of Prym varieties by 
a pair of quadrisecants}, Duke Math.\ J.\ {\bf 152} (2010) 317--371. 

\bibitem{Hirota90}
J.\ Hietarinta and R.\ Hirota: 
{\em Multidromion solutions to the Davey--Stewartson equation},
Phys.\ Lett.\ A {\bf 145} (1990) 237--244.

\bibitem{FoSan}
A.\ S.\ Fokas and P.\ M.\ Santini:
{\em Dromions and a boundary value problem for 
the Davey--Stewartson 1 equation}, Physica D {\bf 44} (1990) 99--130.

\bibitem{GN91} 
C.\ R.\ Gilson and J.\ J.\ C.\ Nimmo: 
{\em A direct method for dromion solutions of the Davey--Stewartson 
equations and their asymptotic properties},
Proc.\ R.\ Soc.\ Lond.\ A 
{\bf 435} 
(1991) 339--357.

\bibitem{Boiti95} 
M.\ Boiti, L.\ Martina and F.\ Pempinelli: 
{\em Multidimensional localized solitons},
Chaos, Solitons \& Fractals 
{\bf 5} (1995) 2377--2417.

\bibitem{Vek02}
V.\ E.\ Vekslerchik: 
{\em Functional representation of the Ablowitz--Ladik hierarchy.\ II},
J.\ Nonlinear Math.\ Phys.\ {\bf 9} (2002) 157--180.

\bibitem{Sada}
T.\ Sadakane:
{\em Ablowitz--Ladik hierarchy and two-component Toda
lattice hierarchy}, 
J.\ Phys.\ A:\ Math.\ Gen.\ {\bf 36} (2003) 87--97.


\end{thebibliography}
\end{document}